\documentclass[lettersize,journal]{IEEEtran}
\IEEEoverridecommandlockouts
\usepackage{amsmath,amsfonts}
\usepackage{booktabs}
\usepackage{algorithmic}
\usepackage{algorithm}
\usepackage{array}
\usepackage[caption=false,font=normalsize,textfont=sf]{subfig}
\usepackage{textcomp}
\usepackage{stfloats}
\usepackage{url}
\usepackage{verbatim}
\usepackage{graphicx}
\usepackage{cite}
\usepackage{multicol}
\usepackage{amssymb}
\usepackage{lineno}
\usepackage{makecell}
\usepackage{indentfirst}
\usepackage{bm}
\usepackage{color}
\usepackage{cite}
\usepackage{epstopdf}
\usepackage{amssymb}
\usepackage{xcolor}
\usepackage{tikz}
\usepackage{mathtools}
\usepackage{pifont}
\usepackage{marvosym}
\usepackage{newtxmath}
\usepackage[colorlinks,
linkcolor=blue,
anchorcolor=blue,
citecolor=blue]{hyperref}

\newcommand{\diag}{\mathrm{diag}}

\newcommand{\tr}{\mathrm{tr}}
\newcommand{\HH}{\mathrm{H}}	
\newcommand{\TT}{\mathrm{T}}

\newtheorem{remark}{\textbf{Remark}}

\newtheorem{proposition}{\textbf{Proposition}}

\newtheorem{lemma}{\textbf{Lemma}}

\begin{document}

\title{\huge Secure Communication in MIMOME Movable-Antenna Systems with Statistical Eavesdropper CSI}

\author{Lei Xie,~\IEEEmembership{Member,~IEEE},  Peilan Wang,~\IEEEmembership{Member,~IEEE}, Guanxiong Shen,~\IEEEmembership{Member,~IEEE},\\ Guyue Li,~\IEEEmembership{Member,~IEEE}, Weidong Mei,~\IEEEmembership{Member,~IEEE}, and Liquan Chen,~\IEEEmembership{Senior Member,~IEEE}
	%\thanks{This work was supported in part by the National Natural Science Foundation	of China under Grants 62501143, and in part by Basic Research Program of Jiangsu under Grants BK20251332.}
	\thanks{L. Xie, G. Shen, G. Li, and L. Chen are with School of Cyber Science and Engineering, Southeast University, Nanjing 210096, China (e-mail: \{leixie,gxshen,guyuelee,lqchen\}@seu.edu.cn).}
	\thanks{P. Wang and W. Mei are with the National Key Laboratory of Wireless Communications, University of Electronic Science and Technology of China, Chengdu 611731, China (e-mail: \{peilan\_wangle,wmei\}@uestc.edu.cn).}
}

\maketitle
	
	\begin{abstract}
		This paper investigates the potential of movable antennas (MAs) to enhance physical layer security within a multiple-input multiple-output multiple-antenna eavesdropper (MIMOME) system. We consider a practical scenario where the transmitter operates with imperfect eavesdropper channel state information (ECSI), knowing only the instantaneous line-of-sight (LoS) component and the statistical properties of non-line-of-sight (NLoS) component. To rigorously quantify secrecy performance under the ECSI uncertainty, we adopt the ergodic secrecy rate (ESR) as the metric. Since deriving an exact analytical expression for the ESR is intractable, we leverage random matrix theory to derive a deterministic equivalent. This avoids heavy Monte Carlo simulations and also provides explicit insights into the effects of channel spatial statistics on secrecy performance. Building upon the deterministic equivalent, we formulate a joint maximization problem for the transmit precoding matrix and the antenna positions at the legitimate transmitter. To tackle the non-convexity of this optimization problem, we develop a comprehensive alternating optimization framework. Specifically, the precoding matrix is optimized via a majorization-minimization (MM) algorithm, where the gradient is computed by solving an implicit fixed-point equation. For the antenna position optimization, the complexity of the objective function prevents the construction of standard MM surrogate. To this end, we further propose a novel AMSGrad-based surrogate function that relies solely on gradient information. We provide a rigorous theoretical proof that guarantees the convergence of this proposed algorithm despite relaxing the strict majorization conditions. Simulation results validate our theoretical findings and demonstrate that the proposed MA-aided scheme significantly outperforms conventional fixed-position antenna systems in terms of ESR.
	\end{abstract}
	
	
\section{Introduction}
The sixth-generation (6G) wireless networks aim to support a wide range of advanced applications that demand ubiquitous connectivity and ultra-high data rates, such as autonomous vehicles, smart factories, digital twins, and the low‑altitude economy \cite{liu2022integrated,xie2023collaborative,jiang2025network}. While conventional networks are primarily optimized for terrestrial users and sensing targets, 6G systems necessitate ubiquitous coverage extending into the low-altitude domain. However, traditional base stations, which typically rely on fixed-position antenna (FPA) system, suffer from limited spatial degrees of freedom (DoFs) \cite{xie2023networked}. This static architecture creates a bottleneck in meeting the wide coverage requirement of 6G. 
 
To address the limitations of FPAs, movable antenna (MA) systems have recently been proposed. By dynamically adjusting antenna elements within a constrained region, MAs enable the transmitter to proactively reshape the channel response, which can enhance communication rate performance  \cite{zhu2023modeling,ma2023mimo,mei2024movable2,wang2025movable}. In particular, \cite{zhu2023modeling} developed a new field-response channel model for MA systems and demonstrated the effectiveness of MAs in boosting the achievable rate performance in a single-MA system. The authors of \cite{ma2023mimo} examined the performance of an MA-enhanced multiple-input multiple-output (MIMO) system. The results indicated that by strategically optimizing antenna positions, an MA array could achieve an approximate 38.1$\%$ improvement in MIMO channel capacity compared to FPA systems. Furthermore, the authors of \cite{mei2024movable2} developed a graph-based algorithm for optimizing MA positions in a single-user multiple-input single-output (MISO) system by discretizing the antenna movement region into a finite set of candidate locations. The proposed method achieves global optimality with polynomial-time complexity. Beyond multipath channel reshaping, MAs have also been shown to outperform FPAs in terms of wide-beam coverage under line-of-sight (LoS) propagation conditions in \cite{wang2025movable}.

Meanwhile, the ubiquitous coverage envisioned for 6G introduces a critical trade-off: the broadcast nature required for communication and sensing functionalities inevitably increases the probability of signal interception by unauthorized eavesdroppers \cite{wang2018survey,xie2023sensing,wei2022toward}. Therefore, physical-layer security (PLS) has become an essential area of 6G networks \cite{kihero20236g,10608156,11004012}.
Unlike upper-layer encryption that relies on computational complexity, PLS harnesses channel condition discrepancies to suppress eavesdropping, boost wireless system secrecy capacity, and avoid security risks induced by the computational constraints of traditional encryption methods. Common PLS techniques, such as artificial noise (AN) injection \cite{5306434,6094170}, cooperative jamming \cite{zheng2010optimal,hu2017cooperative}, and secure beamforming \cite{yang2012cooperative,asaad2022secure}, rely on degrading the signal quality at potential eavesdroppers, which in turn imposes significant demands on the spatial DoFs of the transmitter.

MA system holds considerable promise for enhancing the secrecy capacity of next-generation wireless networks  \cite{cheng2024enabling,mei2024movable,tang2024secure,shen2025movable}. 
Specifically, MAs leverage extra spatial DoFs by antenna movement to enhance the secrecy performance of wireless networks. 
For example, MAs can proactively direct spatial nulls toward eavesdroppers while avoiding deep fading at legitimate receivers. Recent studies have demonstrated the effectiveness of integrating MAs into PLS via tailored antenna position optimization methods. In \cite{cheng2024enabling}, MAs were incorporated into PLS by jointly optimizing antenna positions and beamforming, resulting in significant gains in both secrecy rate and power efficiency compared with conventional FPA systems. In \cite{mei2024movable}, a discrete sampling method was introduced to maximize the secrecy rate in MA-enhanced MISO systems, recasting the continuous antenna placement problem as a discrete point selection task solvable via partial enumeration and sequential update algorithms. Furthermore, \cite{tang2024secure} maximized the secrecy rate of MA-aided MIMO systems through joint optimization of transmit precoding, artificial noise, and antenna locations, achieving substantial security gains over traditional FPA-based designs. 
The authors of \cite{shen2025movable} investigated the MA position optimization problem for concurrent secrecy and multicast information transmission and characterized the associated secrecy capacity region. 
Ultimately, extending conventional PLS frameworks for practical MA systems remains challenging, since secrecy transmit beamforming fundamentally relies on accurate eavesdropper channel state information (ECSI) \cite{li2019beamforming}. However, prior works mostly assume full knowledge of instantaneous ECSI to facilitate tractable analysis and optimization; however, as eavesdroppers are typically passive and non-cooperative, acquiring perfect ECSI is generally difficult in practice.

To mitigate this limitation, there have been a handful of studies that explored more robust system designs that dispense with full ECSI. For instance, \cite{feng2024movable} proposed a virtual MA-based robust framework for scenarios with fully unknown ECSI, where ECSI uncertainty is modeled via the eavesdropper's feasible region to achieve global suppression of signal leakage. Other works modeled the eavesdropper link as a deterministic line-of-sight (LoS) path based on the prior knowledge of eavesdroppers' locations  \cite{hu2024secure,11214460}, which are obtained by pre-transmission sensing \cite{su2023sensing}. However, such deterministic models overlook stochastic multipath and non-line-of-sight (NLoS) components prevalent in dense urban environments, potentially leading to over-optimistic secrecy performance evaluations. To bridge this gap, \cite{hu2024movable} proposed a hybrid ECSI framework for multiple-input single-output multiple-eavesdropper (MISOME) systems, in which the LoS component is perfectly known while the NLoS component is statistically known.
However, to the best of our knowledge, there is no existing works focusing on the robust system designs for more general multiple-input multiple-output multiple-antenna eavesdropper (MIMOME) scenarios, while they are practically relevant considering the fact that modern eavesdroppers can easily deploy multiple antennas to enhance their intercepting capability. However, due to the matrix-valued nature of the MIMOME channels, the existing algorithms designed for MISOME systems cannot be directly applied.

To fill in this gap, this paper investigates the robust performance optimization in an MIMOME-MA system, where multiple antennas are deployed at both the legitimate user and the eavesdropper. 
In particular, we focus on two critical and open research questions: 1) How to evaluate the secrecy performance with statistic ECSI; and 2) How to optimize the secrecy performance through joint precoding and antenna position designs.
To answer these questions, we model the ECSI as the superposition of a deterministic LoS component and a random NLoS component and derive the ergodic secrecy rate (ESR) in terms of the random NLoS component. As the ESR expression is intractable to handle, we further leverage random matrix theory (RMT) to derive a closed-form determinstic equivalent of the exact ESR. Subsequently, we formulate a joint optimization problem for the precoding matrix and antenna positions to maximize the ESR, which is solved via the proposed alternating optimization (AO) framework. Simulation results validate the accuracy of the theoretical analysis results and the effectiveness of the proposed optimization method.

	The main contributions of this paper are summarized as:
	\begin{enumerate}
		\item We derive an explicit deterministic equivalent for the ESR in the MIMOME-MA system with only statistical ECSI. By leveraging RMT tools, we can bypass the computational prohibitiveness of traditional Monte Carlo simulations. This theoretical result not only captures the asymptotic behavior of the system, but also provides an explicit and differentiable objective function, which will serve as the foundation for the subsequent joint optimization of the precoding matrix and antenna positions.
	    \item For the transmit precoding design, we develop an algorithm based on the majorization-minimization (MM) framework. In particular, we address the non-convexity of the problem by constructing a sequence of tractable convex surrogate problems that locally approximate the original objective. A key technical contribution in this step is the derivation of the gradient for the ESR, which necessitates solving an implicit fixed-point equation inherent to the deterministic equivalent.
		\item The continuous position optimization of MAs poses a significant challenge. In particular, it is hard to construct a standard MM surrogate function (i.e., one that strictly satisfies tangent and lower-bound conditions), due to the complicated expression of the ESR with respect to (w.r.t.) antenna position. To address this challenge, we propose an AMSGrad-based surrogate function that relies solely on gradient information. Although this approach relaxes the strict majorization requirements, we provide a rigorous proof guaranteeing that the proposed algorithm converges to a stationary point. 	
	\end{enumerate}
		
	The remainder of this paper is organized as follows. Sec. II describes the system model and outlines the fundamental assumptions regarding channel state information.  Sec. III provides a detailed derivation of the explicit expression for the ESR. 
	Sec. IV formulates the system design problem, and then proposes a framework for joint precoding and antenna position optimization to maximize the ESR. Sec. V presents numerical results to validate our theoretical analysis and demonstrate the superiority of the proposed scheme compared to benchmarks. Finally, Sec. VI concludes the paper.
	
	\emph{Notations:}
Matrices and vectors are denoted by bold upper and lower case letters, e.g., $\mathbf{A}$ and $\mathbf{a}$, respectively, while $[\mathbf{A}]_{i,j}$ denotes the $(i,j)$th element of matrix $\mathbf{A}$. $\mathbb{C}^{M\times N}$ and $\mathbb{R}^{M\times N}$ denote the space of $M\times N$ complex and real matrices, respectively. $\mathbb{E}(\cdot)$ denotes the expectation operation. $||\mathbf{x}||_2$ denotes the 2-norm of vector $\mathbf{x}$.  $|\mathbf{x}|^2$ denotes the element-wise power. $||\mathbf{A}||_2$ and $||\mathbf{A}||_F$ denote the spectral norm and the Frobenius norm of matrix $\mathbf{A}$, respectively. $\tr(\mathbf{A})$ and $|\mathbf{A}|$ denote the trace operation and the determinant of matrix $\mathbf{A}$, respectively. $\bm{\nabla}_\mathbf{x} f(\mathbf{x})$ and $\bm{\nabla}_\mathbf{x}^2 f(\mathbf{x})$ denote the gradient vector and the Hessian matrix of the function $f$ w.r.t. the vector $\mathbf{x}$. $\mathcal{CN}(\mathbf{0},\mathbf{R})$ denotes the circularly symmetric complex Gaussian distribution with zero mean and covariance matrix $\mathbf{R}$. $\diag{\cdot}$ denotes the diagonalization operation. $(\cdot)^\TT$ and $(\cdot)^\HH$ denote the transpose and the Hermitian transpose, respectively. For a complex value $x$, $\Re(x)$ and $\Im(x)$ denote the real and imaginary parts of $x$, respectively. $[x]^+$ denotes the operator to the non-negative value, i.e., $[x]^+ \triangleq \max\{0,x\}$.

	\section{System Model}	
	\begin{figure}[t]
		\centering
		\includegraphics[width=3.6in]{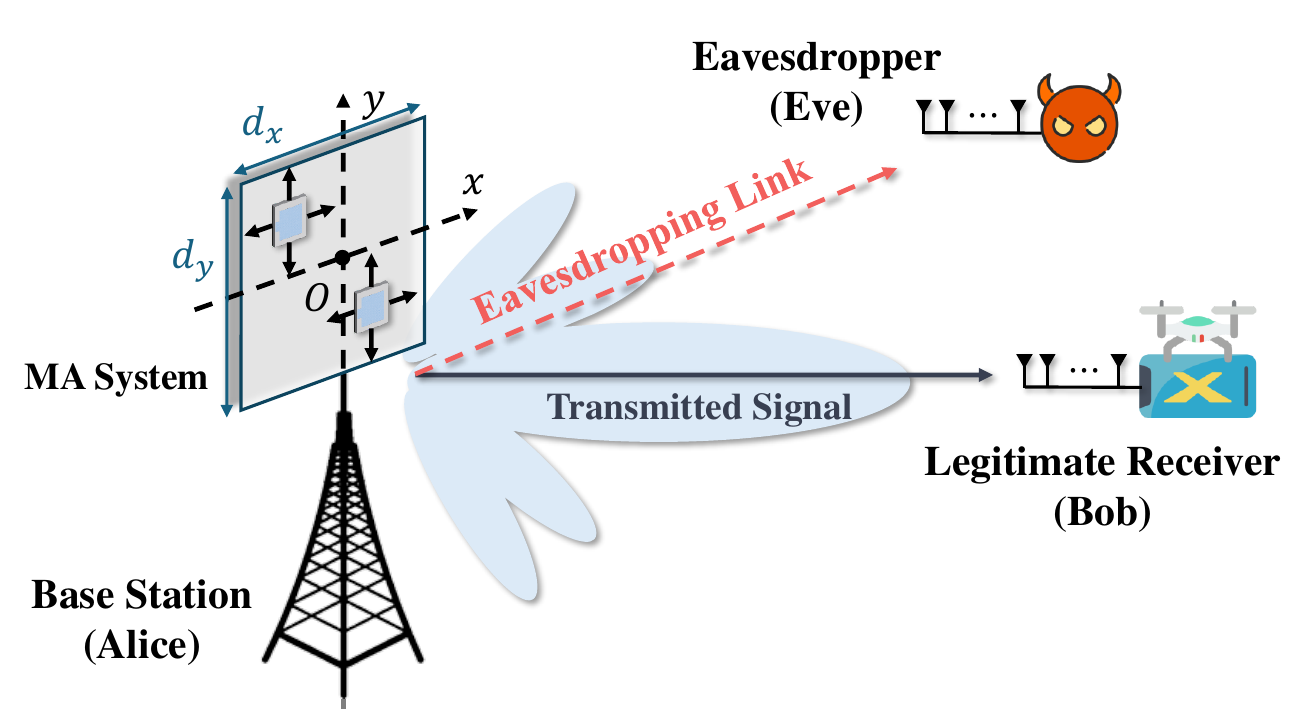}
		\caption{Illustration of the considered MIMOME system.}
		\label{fig_sysill}
	\end{figure}
	Consider the downlink of a secure MIMOME transmission system, which consists of a base station (Alice) and a legitimate receiver ($\mathrm{Bob}$), as illustrated in Fig. \ref{fig_sysill}. The base station is equipped with $N$ MAs, which are deployed in a two-dimensional (2D) planar region within the $x$-$y$ plane. Unlike conventional FPA arrays, the MAs considered in this paper are capable of continuous movement to exploit additional spatial degrees of freedom. Specifically, each antenna element is constrained to move within a rectangular region of dimensions $D_x \times D_y$, where $D_x$ and $D_y$ denote the maximum movement ranges along the $x$-axis and $y$-axis, respectively. Mathematically, we have
	\begin{equation}
		-\mathbf{t}_{\max} \leq \mathbf{t}_n \leq \mathbf{t}_{\max},
	\end{equation}
	where %$\mathbf{t}_{\max} = [D_x,D_y]^\TT$.
	\begin{equation}
		\mathbf{t}_{\max} = \left[\begin{matrix}
			D_x\\
			D_y
		\end{matrix}\right].
	\end{equation}
	Bob is equipped with uniform linear array with $L$ antennas. The BS transmits $M$ data stream to Bob ($M\leq N,L$). 
	There are $M$ single-antenna eavesdroppers ($\mathrm{Eve}_{m},m=1,\cdots,M$) which can cooperate with each other, and thus form a virtual $M$‑antenna receiver (Eve). 
	In the following, we will introduce the channel model and the received signal, respectively. 
	
	\subsection{Channel Model}	
	In practical scenarios, the LoS path is generally superimposed on NLoS paths arising from reflections off the ground and other nearby objects. 
	
	\subsubsection{Eve's channel}
\begin{figure}[t]
\centering
\includegraphics[width=2.6in]{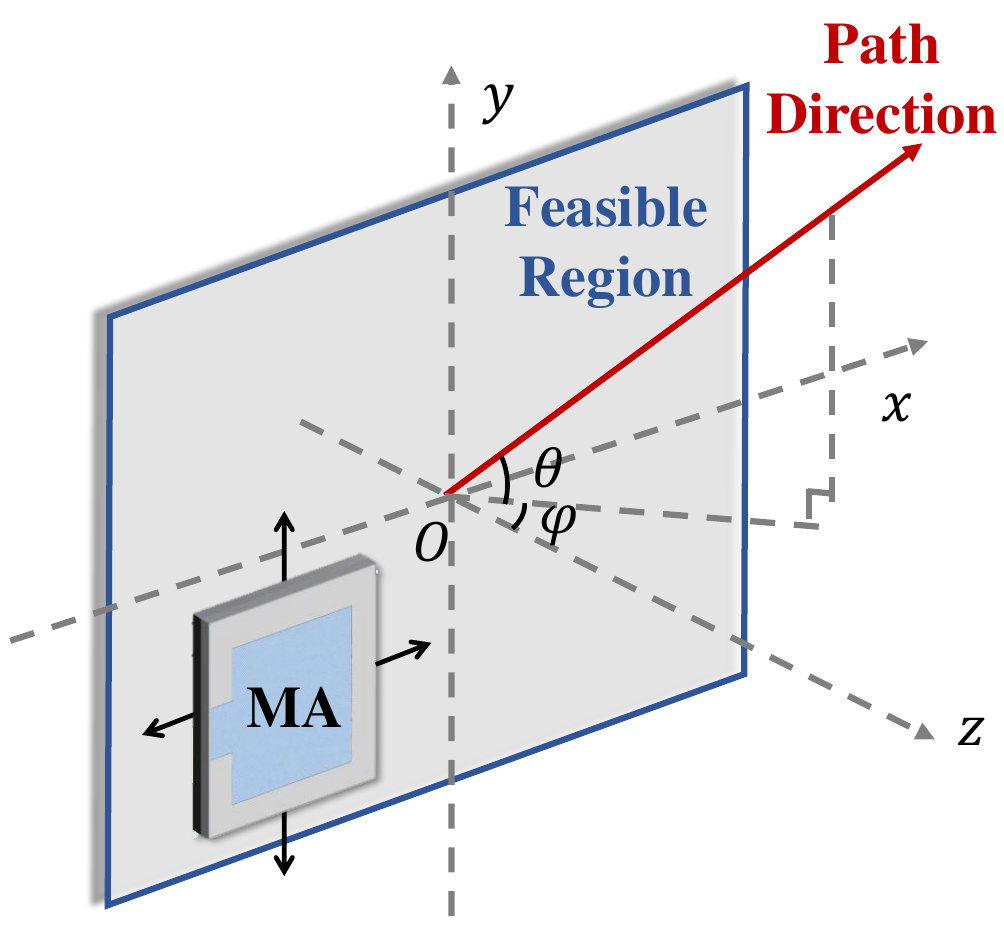}
\caption{Illustration of the coordinates and spatial angles for the MA.}
\label{fig_sysill2}
\end{figure}

	Considering the covert nature of adversarial attacks, Eve is often located in a cluttered environment where the direct link to the transmitter is partially or fully occluded by vegetation or buildings. This shadowing effect implies that the LoS component in the Eve’s channel may be significantly more attenuated than that of the legitimate user. To accurately capture this discrepancy in multipath propagation, we model Eve’s channel as Rician fading, which provides the flexibility to represent various scattering conditions by adjusting the ratio between the LoS and NLoS components. In particular, the channel vectors between Alice and $\mathrm{Eve}_m$ is denoted by \cite{kang2006capacity}
	\begin{equation}
		\mathbf{g}_m = \sqrt{\frac{K_{e,m} \beta_{e,m}}{K_{e,m}+1}} \mathbf{g}_m^{\mathrm{LoS}}(\mathbf{T}) + \sqrt{\frac{\beta_{e,m}}{K_{e,m}+1}}\mathbf{g}_m^{\mathrm{NLoS}},
	\end{equation}
	where $K_{e,m}$ and $\beta_{e,m}$ denote the Rician K-factor and the path loss corresponding to $\mathrm{Eve}_m$, and the set of the antenna positions $\mathbf{T}$ is given by
	\begin{equation}
		\mathbf{T} = [\mathbf{t}_1,\mathbf{t}_2,\cdots,\mathbf{t}_N]. 
	\end{equation}
	In particular, the LoS deterministic component is given by
	\begin{equation}
		\begin{split}
		\mathbf{g}_m^{\mathrm{LoS}}(\mathbf{T}) =  \mathbf{a}(\mathbf{T},\bm{\rho}_{e,m}),
		\end{split}
	\end{equation}
	where $\mathbf{a}(\mathbf{T},\bm{\rho}_{e,m})$ denotes the field-response vector, i.e., 
	\begin{equation}\label{aitrho}
		\begin{split}
			\mathbf{a}(\mathbf{T},\bm{\rho}_{e,m}) = \left[e^{j\frac{2\pi}{\lambda}\mathbf{t}_1^\TT\bm{\rho}_{e,m}},e^{j\frac{2\pi}{\lambda}\mathbf{t}_2^\TT\bm{\rho}_{e,m}},\cdots,e^{j\frac{2\pi}{\lambda}\mathbf{t}_N^\TT\bm{\rho}_{e,m}}\right]^\TT.
		\end{split}
	\end{equation}
	In \eqref{aitrho}, $\bm{\rho}_{e,m} = \left[\sin\theta_{e,m}\cos\varphi_{e,m},\cos\theta_{e,m}\right]^\TT$, where $\theta_{e,m}$ and $\varphi_{e,m}$ are the elevation and azimuth angles of the LoS path corresponding to $\mathrm{Eve}_m$, respectively, as illustrated in Fig. \ref{fig_sysill2}. The wavelength is denoted by  $\lambda$. The NLoS component $\mathbf{g}_m^{\mathrm{NLoS}}(\mathbf{T})$ consists of i.i.d. small‑scale fading entries, each following $\mathcal{CN}(0,1)$.
	
	\subsubsection{Bob's channel}
	Unlike passive eavesdroppers, Bob is an authorized communication partner and therefore does not require covert transmission. The channel between Alice and Bob is assumed to be perfectly known, which is given by \cite{ma2023mimo}
	\begin{equation}
		\mathbf{H} = \mathbf{A}_T \mathbf{\Lambda}_b \mathbf{A}_R^\HH,
	\end{equation}
	where $\mathbf{A}_T = \left[ \mathbf{a}(\mathbf{T},\bm{\rho}_{b,1}),\mathbf{a}(\mathbf{T},\bm{\rho}_{b,2}),\cdots,\mathbf{a}(\mathbf{T},\bm{\rho}_{b,L}) \right]$,  $\mathbf{\Lambda}_b = \diag \left(\beta_{b,l},l=1,2,\cdots,L\right)$, and $\mathbf{A}_R = [\mathbf{b}_1,\mathbf{b}_2,\cdots,\mathbf{b}_L]$.	In particular, $\beta_{b,l}$ denote the path loss corresponding to the $l$th path, $\bm{\rho}_{b,l} = \left[\sin\theta_{b,l}\cos\varphi_{b,l},\cos\theta_{b,l}\right]^\TT$, where $\theta_{b,l}$ and $\varphi_{b,l}$ are the elevation and azimuth angles of the $l$th path at Alice, and $\mathbf{b}_l$ denotes the field-response vector of the $l$th path.

	\subsection{Received Signal Model}
	The received signal at $\mathrm{Bob}$ is given by
	\begin{equation}
		\begin{split}
			\mathbf{y}_b = \mathbf{H}^\HH \mathbf{F} \mathbf{s} + \mathbf{n}_b, 
		\end{split}
	\end{equation}
	where $\mathbf{F}\in\mathbb{C}^{N\times M}$ denotes the precoding vector, $\mathbf{s}$ denotes the transmitted signal, and $\mathbf{n}_b$ denotes the additive white Gaussian noise with zero mean and covariance $\sigma^2 \mathbf{I}$. In particular, the transmitted symbol $\mathbf{s}$ are independent and identically distributed (i.i.d.)  drawn from a standard complex constellation, whose $k$th entry $s_k$ satisfies 
	\begin{equation}
	\mathbb{E}[|s_k|^2]=1,\;\mathbb{E}[s_k]=0,\; \mathbb{E}[s_k^2]=0.
	\end{equation}
	
	Since the eavesdroppers can cooperate with each other, they effectively form a multi‑antenna receiver. 
	Similarly, the received signal at the effective Eve is given by
 	\begin{equation}
	 	\begin{split}
	 		\mathbf{y}_e = \mathbf{G}^\HH \mathbf{F} \mathbf{s} + \mathbf{n}_e, 
	 	\end{split}
	 \end{equation}
	where $\mathbf{G} = [\mathbf{g}_1,\mathbf{g}_2,\cdots, \mathbf{g}_M] \in\mathbb{C}^{N\times M}$ denotes the eavesdropper channel, and $\mathbf{n}_e$ denotes the additive white Gaussian noise with zero mean and covariance $\sigma^2 \mathbf{I}$. Specifically, an eavesdropper with $M$ antennas can be viewed as a special case of our model where $M$ single-antenna users are co-located.

	\begin{remark}
	Due to the passive nature of eavesdroppers, the instantaneous ECSI is typically unavailable. Consequently, this paper proceeds under the assumption that the transmitter knows perfect CSI for the legitimate receiver (Bob), while only knowing the statistical ECSI \cite{hu2024movable}. This assumption is practical and well-justified in scenarios where potential eavesdroppers were previously active, legitimate users of the network. In such contexts, the transmitter can leverage historical data regarding their coarse locations and large-scale fading characteristics to estimate statistical channel properties.
	\end{remark}

\section{Ergodic Secrecy Rate}
	Typically, secrecy performance is quantified by the secrecy rate, which characterizes the information rate that can be delivered reliably to the legitimate receiver while remaining confidential from Eves. In particular, the secrecy rate of the considered system is given by \cite{asaad2022secure}
	\begin{equation}
		\begin{split}
			\mathcal{R}(\mathbf{F},\mathbf{T}) = \left[\mathcal{R}_{b} (\mathbf{F},\mathbf{T}) - \mathcal{R}_{e} (\mathbf{F},\mathbf{T})\right]^+.
		\end{split}
	\end{equation}
	In particular, $\mathcal{R}_{b} (\mathbf{F},\mathbf{T})$ denotes the achievable rate to $\mathrm{Bob}$, which is given by
\begin{equation}
	\mathcal{R}_{b} (\mathbf{F},\mathbf{T}) = \log\left\vert\mathbf{I}+\sigma^{-2}\mathbf{H}^\HH\mathbf{F}\mathbf{F}^\HH\mathbf{H}\right\vert.
\end{equation}

	 For a given realization of \(\{\boldsymbol{\beta}_{e,m}\}_{m=1}^M\), the capacity of eavesdroppers is given by
	\begin{equation}
		\begin{split}
				\mathcal{R}_{e} \left(\mathbf{F},\mathbf{T}\right) =  \log\left\vert\mathbf{I}+\sigma^{-2}\mathbf{G}^\HH\mathbf{F}\mathbf{F}^\HH\mathbf{G}\right\vert.
			\end{split}
	\end{equation}

\subsection{ESR with Statistic ECSI}
	Owing to the unavailability of instantaneous ECSI, the ESR is adopted as the performance metric.  Specifically, the ESR at Bob is defined as
	\begin{equation}
		\overline{\mathcal{R}}(\mathbf{F},\mathbf{T}) = \left[ \mathcal{R}_{b}(\mathbf{F},\mathbf{T}) - \overline{\mathcal{R}}_{e}(\mathbf{F},\mathbf{T}) \right]^+,
	\end{equation}
	where the average achievable rate to Eve is given by
	\begin{equation}
		\overline{\mathcal{R}}_{e}(\mathbf{F},\mathbf{T}) = \mathbb{E}_{\mathbf{G}} \left[ \mathcal{R}_{e}(\mathbf{F},\mathbf{T}) \right].
	\end{equation}
	Here, the expectation $\mathbb{E}_{\mathbf{G}}[\cdot]$ is taken over the small-scale fading distribution of the NLoS components of Eve's channel. 
	
	This metric effectively characterizes the long-term average secrecy performance when the transmitter only has access to the statistical ECSI. However, the presence of the expectation operator $\mathbb{E}_{\mathbf{G}}[\cdot]$ poses a significant challenge to the optimization process, as it inherently complicates the derivation of a closed-form or even a tractable analytical expression for the objective function. Consequently, standard optimization techniques cannot be directly applied, necessitating a more sophisticated approach to handle the average rate term.

	\subsection{Asymptotic Approximation of ESR}
	To derive a tractable expression for the ESR under the statistical ECSI, we state the following proposition.

\begin{proposition}
	\label{prop:ergodic_secrecyC}
	Define the matrix $\mathbf{B}$ as
	\begin{equation}\label{bk}
		\mathbf{B} =  \mathbf{\Lambda}^{\frac{1}{2}}\mathbf{G}_{\mathrm{LoS}}^\HH\mathbf{F}.
	\end{equation}
	where $\mathbf{G}_{\mathrm{LoS}} = \left[\mathbf{g}_1^{\mathrm{LoS}}(\mathbf{T}),\mathbf{g}_2^{\mathrm{LoS}}(\mathbf{T}),\cdots,\mathbf{g}_M^{\mathrm{LoS}}(\mathbf{T})\right]$, and
	\begin{equation}
	\mathbf{\Lambda} = \diag\left({\frac{K_{e,m} \beta_{e,m}}{K_{e,m}+1}},m=1,2,\cdots,M\right).	
	\end{equation}
	Define two diagonal matrices $\mathbf{D}$ and $\widetilde{\mathbf{D}}$ as
		\begin{equation}\label{D_def_0}
			\mathbf{D} = \diag\left(\frac{M\beta_{e,m} }{K_{e,m}+1},m=1,2,\cdots,M\right),
		\end{equation}
		\begin{equation}\label{Dtilde_def_0}
			\widetilde{\mathbf{D}} = \diag\left(||\mathbf{f}_n||^2,n=1,2,\cdots,M\right).
		\end{equation}
		
		Then, define $(\delta,\tilde{\delta})$ as the unique positive solution of the following fixed-point equation system:
	\begin{equation}\label{fpequ_0}
		\begin{split}
			\left\{
			\begin{array}{l}
				\delta = \frac{1}{M} \tr\left(\mathbf{D}\mathbf{\Gamma}\right)\\
				\tilde{\delta} = \frac{1}{M}\tr\left(\widetilde{\mathbf{D}}\widetilde{\mathbf{\Gamma}}\right)\\
			\end{array}\right.,
		\end{split}
	\end{equation}
	where
	\begin{equation}
		\begin{split}
			\left\{
			\begin{array}{l}
				\mathbf{\Gamma}=
				\left[\sigma^2 \mathbf{\Phi}^{-1}+\mathbf{B}\widetilde{\mathbf{\Phi}}\mathbf{B}^\HH\right]^{-1}\\ 
				\widetilde{\mathbf{\Gamma}}=\left[\sigma^2 	\widetilde{\mathbf{\Phi}}^{-1}+\mathbf{B}^\HH\mathbf{\Phi}\mathbf{B}\right]^{-1}\\
				\mathbf{\Phi} = \left(\mathbf{I}+\tilde{\delta}\mathbf{D}\right)^{-1}\\
				\widetilde{\mathbf{\Phi}} = \left(\mathbf{I}+\delta\widetilde{\mathbf{D}}\right)^{-1}\\
			\end{array}\right..
		\end{split}
	\end{equation}
As $N,M \to \infty$, the ergodic secrecy rate can be given by
\begin{equation}\label{App1}
\begin{split}
	\overline{\mathcal{R}}_{e}\overset{a.s.}{\to}-\log \left\vert\sigma^{2}\mathbf{\Gamma}\right\vert + \log \left\vert\mathbf{I}+\delta\widetilde{\mathbf{D}}\right\vert - \sigma^2 M \delta \tilde{\delta}. 
\end{split}
\end{equation}
\end{proposition}
	
\textbf{\emph{Proof}}: 	Note that $\mathcal{R}$ consists of two parts, i.e., $\mathcal{R}_{b}$ and $\mathcal{R}_{e}$. Given the channel to Bob is known, $\mathcal{R}_{b}$ is a deterministic variable. The proof reduces to an asymptotic analysis of $\mathcal{R}_{e}$.
Concretely, we establish a deterministic equivalent for $\mathcal{R}_{e}$ in the large‑system regime and then use this equivalent to obtain the closed‑form expression stated in \textbf{\emph{Proposition \ref{prop:ergodic_secrecyC}}}. The argument follows the approach in random‑matrix theory and extends the result of \cite[Theorem 1]{5429113}, which is summarized below.

\begin{lemma}[{\cite[Theorem 1]{5429113}}]
	\label{lem:det_eq}
	Let $\mathbf{\Sigma} = \mathbf{B}+\mathbf{Y} \in \mathbb{C}^{r\times t}$ and assume $\mathbf{Y} = \frac{1}{\sqrt{t}} \mathbf{D}^{\frac{1}{2}}\mathbf{X}\widetilde{\mathbf{D}}^{\frac{1}{2}}$, where $\mathbf{D}$ and $\widetilde{\mathbf{D}}$ represent the diagonal matrices $\mathbf{D} = \diag \left(d_i,i=1,2,\cdots,r\right)$ and $\widetilde{\mathbf{D}} = \diag \left(\tilde{d}_j,j=1,2,\cdots,t\right)$. $\mathbf{X}$ is a matrix whose entries are i.i.d. with mean $0$ and variance $1$. Then, $(\delta,\tilde{\delta})$ is defined as the unique positive solution of the following fixed-point equation system:
	\begin{equation}
		\begin{split}
			\left\{
			\begin{array}{l}
				\delta = \frac{1}{t} \tr\left(\mathbf{D}\mathbf{\Gamma}\right)\\
				\tilde{\delta} = \frac{1}{t}\tr\left(\widetilde{\mathbf{D}}\widetilde{\mathbf{\Gamma}}\right)\\
			\end{array}\right.,
		\end{split}
	\end{equation}
	where
	\begin{equation}
		\begin{split}
			\left\{
			\begin{array}{l}
				\mathbf{\Gamma}=
				\left[\sigma^2 \mathbf{\Phi}^{-1}+\mathbf{B}\widetilde{\mathbf{\Phi}}\mathbf{B}^\HH\right]^{-1}\\ 
				\widetilde{\mathbf{\Gamma}}=\left[\sigma^2 \widetilde{\mathbf{\Phi}}^{-1}+\mathbf{B}^\HH\mathbf{\Phi}\mathbf{B}\right]^{-1}\\
				\mathbf{\Phi} = \left(\mathbf{I}+\tilde{\delta}\mathbf{D}\right)^{-1}\\
				\widetilde{\mathbf{\Phi}} = \left(\mathbf{I}+\delta\widetilde{\mathbf{D}}\right)^{-1}\\
			\end{array}\right..
		\end{split}
	\end{equation}

	The following approximation holds true:
	\begin{equation}
		\begin{split}
			\mathbb{E}_{\mathbf{X}}\log\left\vert\mathbf{I}+\sigma^{-2}\mathbf{\Sigma}\mathbf{\Sigma}^\HH\right\vert = \overline{\rho} + \mathcal{O}\left(\frac{1}{t}\right),
		\end{split}
	\end{equation}
	where
	\begin{equation}
		\begin{split}
			\overline{\rho} = - \log \left\vert\sigma^{2}\mathbf{\Gamma}\right\vert + \log \left\vert\mathbf{I}+\delta\widetilde{\mathbf{D}}\right\vert - \sigma^2 t \delta \tilde{\delta}. 
		\end{split}
	\end{equation}
\end{lemma}

The main idea of this proof is to recast $\mathcal{R}_{e}$ into an extended model which fits into the framework of \cite[Theorem 1]{5429113}.

First, we have
\begin{equation}
	\mathcal{R}_{e}= \log \left\vert \mathbf{I}+\frac{1}{\sigma^2}\mathbf{Z}\mathbf{Z}^\HH\right\vert,
\end{equation}
where
\begin{equation}
	\mathbf{Z} =  \mathbf{G}^\HH\mathbf{F}.
\end{equation}
Note that the $(m,n)$th entry of $\mathbf{Z}$ is Gaussian, with mean $[\mathbf{B}]_{m,n}$ and variance given by
\begin{equation}
	\sigma_{z,(m,n)}^2 = \frac{\beta_{e,m}||\mathbf{f}_n|| ^2}{K_{e,m}+1},
\end{equation}
where $\mathbf{f}_n$ denotes the $n$th column of $\mathbf{F}$. 
Let $\mathbf{X}$ be a matrix whose entries are i.i.d. with mean $0$ and variance $1$. Then, $\mathbf{Z}$ is equivalent to
\begin{equation}
	\mathbf{Z} \sim \mathbf{B} + \frac{1}{\sqrt{M}} \mathbf{D}^{\frac{1}{2}}\mathbf{X}\widetilde{\mathbf{D}}^{\frac{1}{2}},
\end{equation}
where $\mathbf{B}$, $\mathbf{D}$, and $\widetilde{\mathbf{D}}$ are defined in \eqref{bk}, \eqref{D_def_0}, and \eqref{Dtilde_def_0}, respectively. 
By applying \emph{\textbf{Lemma \ref{lem:det_eq}}}, we have
\begin{equation}
	\mathbb{E}_{\mathbf{G}}(\mathcal{R}_{e})\overset{a.s.}{\to} -\log \left\vert\sigma^{2}\mathbf{\Gamma}\right\vert + \log \left\vert\mathbf{I}+\delta\widetilde{\mathbf{D}}\right\vert - \sigma^2  M \delta \tilde{\delta},
\end{equation}
where $(\delta,\tilde{\delta})$ is defined as the solution to the fixed-point equation in \eqref{fpequ_0}. \hfill $\blacksquare$
	
\begin{remark}
This proposition provides an analytically-tractable  deterministic equivalent for $\overline{\mathcal{R}}_{e} (\mathbf{F},\mathbf{T})$.
Although \eqref{App1} is derived in the asymptotic regime with $N,M\to \infty$, simulation results in Sec. \ref{Sim_Res} will demonstrate that this approximation remains accurate even in the moderate regime. 
The approximation substantially simplifies optimization by replacing stochastic objectives with deterministic surrogates. 
\end{remark}
	
%	\subsection{ESR}

	\section{Problem Formulation and Solution}
	In this paper, we aim to jointly optimize the precoding $\mathbf{F}$ and the antenna position $\mathbf{T}$ to maximize the ESR. The optimization problem is formulated as
	\begin{equation}\label{Prob_00}
		\begin{split}
			\mathcal{P}_0: \max_{\mathbf{F},\mathbf{T}} \quad & \mathcal{R}_{b} (\mathbf{F},\mathbf{T}) - \overline{\mathcal{R}}_{e} (\mathbf{F},\mathbf{T})\\
			 s.t.\quad \; & (\mathrm{C1}): \tr\left(\mathbf{F}\mathbf{F}^\HH\right) \leq P, \\
			&(\mathrm{C2}): \Vert\mathbf{t}_n-\mathbf{t}_m\Vert^2 \geq d,\quad  m \neq n,\\
			&(\mathrm{C3}): \mathbf{t}_n \in \mathcal{T}, \quad n=1,2,\cdots,N,
			\end{split}
	\end{equation}
	where C1 constrains transmit power to a maximum value of $P$, C2 requires that the distance between any two distinct antennas must be at least the prescribed constant $d$, and C3 restricts each antenna to lie within a prescribed region $\mathcal{T} = \{\mathbf{t}_n | 	-\mathbf{t}_{\max} \leq \mathbf{t}_n \leq \mathbf{t}_{\max}\}$.

	Given that problem \eqref{Prob_00} involves two variables, we propose to solve this problem based on the AO framework \cite{xie2022perceptive}. The resulting AO framework at the $k$th iteration for \eqref{Prob_00} proceeds as follows:
	\begin{align}
		\mathbf{F}_{k+1} &= \arg \max_{\mathbf{F}\in \mathcal{S}_{F}} \mathcal{R}_{b} (\mathbf{F};\mathbf{T}_{k}) - \overline{\mathcal{R}}_{e} (\mathbf{F};\mathbf{T}_{k}),\label{prob_221}\\
		\mathbf{T}_{k+1} &= \arg \max_{\mathbf{T}\in \mathcal{S}_{T}} \mathcal{R}_{b} (\mathbf{T};\mathbf{F}_{k+1}) - \overline{\mathcal{R}}_{e} (\mathbf{T};\mathbf{F}_{k+1}),\label{prob_222}
	\end{align}
	where 
\begin{equation}	
	\mathcal{S}_{F} = \left\{\mathbf{F}|\tr\left(\mathbf{F}\mathbf{F}^\HH\right) \leq P\right\},
\end{equation}
\begin{equation}
\mathcal{S}_{T} = \left\{\mathbf{t}_n|\{\Vert\mathbf{t}_n-\mathbf{t}_m\Vert^2 \geq d\}_{m \neq n},\; \{\mathbf{t}_n \in \mathcal{T}\}_{n=1}^N\right\}.
\end{equation}
These steps are repeated until convergence, which produce a stationary point of problem $\mathcal{P}_0$  \cite{xie2022perceptive}.

	\subsection{Update $\mathbf{F}$ with Fixed $\mathbf{T}_{k}$}
	Typically, problem \eqref{Prob_00} can be solved by using the interior-point method in conjunction with Newton’s method, a process demanding the derivation of both the gradient and the Hessian matrix. Nevertheless, given that $(\delta,\tilde{\delta})$ is obtained by solving the equation defined in \eqref{fpequ_0}, the analytical computation of the Hessian matrix for $\overline{\mathcal{R}}_e\left(\mathbf{F},\mathbf{T}\right)$ presents significant difficulty. To mitigate this issue, we propose a gradient projection (GP)-based method, which simplifies the optimization by effectively handling constraints and only necessitating the gradient computation.

	Due to the concavity nature of logarithmic function, $\mathcal{R}_{b}$ is convex in $\mathbf{F}$, while $-\overline{\mathcal{R}}_{e}$ is concave in $\mathbf{F}$. Therefore, the overall objective function $\mathcal{L}$ is non-convex. 
	To handle this, we adopt the MM framework, which transforms the original problem into a sequence of simpler subproblems \cite{sun2016majorization}. At the $t$th  iteration, we optimize a surrogate function that lower‑bounds the objective $\mathcal{G}\left(\mathbf{F};\mathbf{F}_{k,t}\right)$. Before proceeding, we give the gradient of $\overline{\mathcal{R}}_e\left(\mathbf{F},\mathbf{T}\right)$ with respect to $\mathbf{F}$ in the following proposition.

		\begin{proposition}\label{grad_fk}
		Define $\hat{\delta}(i,j)$ as the Kronecker delta, i.e., 
		\begin{equation}
			\begin{split}
				\hat{\delta}(i,j) = 
				\begin{cases}
					1, & \text{if } i = j \\
					0, & \text{if } i \neq j
				\end{cases}.
			\end{split}
		\end{equation}
		By differentiating the equations in \eqref{fpequ_0} w.r.t. the $(i,j)$th entry of $\mathbf{F}$, denoted by $F_{(i,j)}$, we establish the following fixed-point equations.
		\begin{equation}\label{fpequ_grad_1}
			\begin{split}
				\left\{
				\begin{array}{l}
					\delta_{(i,j)}'= \frac{1}{M} \tr \left(\mathbf{D}\mathbf{\Gamma}_{(i,j)}'\right)\\
					\tilde{\delta}_{(i,j)}'= \frac{1}{M}\tr\left(\widetilde{\mathbf{D}}\widetilde{\mathbf{\Gamma}}_{(i,j)}'\right) + \frac{1}{M} F_{(i,j)} \hat{\delta}(i,j)\\
				\end{array}\right.,
			\end{split}
		\end{equation}
		where
		\begin{subequations}
			\begin{equation}
				\begin{split}
					\mathbf{\Xi}_{(i,j)}'
					& =\mathbf{B}\widetilde{\mathbf{\Phi}}\mathbf{e}_j\mathbf{e}_i^\TT	\mathbf{G}_{\mathrm{LoS}}\mathbf{\Lambda}^{\frac{1}{2}}+\mathbf{B}	\widetilde{\mathbf{\Phi}}_{i,j}'\mathbf{B}^\HH,
				\end{split}
			\end{equation}
			\begin{equation}
				\begin{split}
					\widetilde{\mathbf{\Xi}}_{(i,j)}' =&\mathbf{e}_j\mathbf{e}_i^\TT \mathbf{G}_{\mathrm{LoS}}\mathbf{\Lambda}^{\frac{1}{2}}\mathbf{\Phi} \mathbf{B}+\mathbf{B}^\HH \mathbf{\Phi}_{(i,j)}'\mathbf{B}.
				\end{split}
			\end{equation}
			\begin{equation}
				\begin{split}	\mathbf{\Gamma}_{(i,j)}'= - \mathbf{\Gamma}\left(\sigma^2\tilde{\delta}_{(i,j)}' \mathbf{D}+ \mathbf{\Xi}_{(i,j)}' \right)\mathbf{\Gamma},
				\end{split}
			\end{equation}
			\begin{equation}
				\begin{split}
					\widetilde{\mathbf{\Gamma}}_{(i,j)}'= - \widetilde{\mathbf{\Gamma}} \left(\sigma^2 \delta F_{(i,j)}\hat{\delta}(i,j) + \sigma^2\delta_{(i,j)}'\widetilde{\mathbf{D}}+ \widetilde{\mathbf{\Xi}}_{(i,j)}' \right)\widetilde{\mathbf{\Gamma}},
				\end{split}
			\end{equation}
			\begin{equation}
				\begin{split}
					\mathbf{\Phi}_{(i,j)}' =  - \mathbf{\Phi} \left( \tilde{\delta}_{(i,j)}' \mathbf{D} \right) \mathbf{\Phi},
				\end{split}
			\end{equation}
			\begin{equation}
				\begin{split}
					\widetilde{\mathbf{\Phi}}_{(i,j)}' = -\widetilde{\mathbf{\Phi}} \left(\delta F_{(i,j)}\hat{\delta}(i,j) + \delta_{(i,j)}'\widetilde{\mathbf{D}}\right)\widetilde{\mathbf{\Phi}}.
				\end{split}
			\end{equation}
		\end{subequations}
		The gradient of $\overline{\mathcal{R}}_e\left(\mathbf{F},\mathbf{T}\right)$ w.r.t. $\mathbf{F}$ is denoted by $\bm{\nabla}_{e,\mathbf{F}} \triangleq \frac{\partial \overline{\mathcal{R}}_e\left(\mathbf{F},\mathbf{T}\right)}{\partial \mathbf{F}}$, whose $(i,j)$th entry is given by
		\begin{equation}\label{Delta_Omega}
			\begin{split}
				&\left[\bm{\nabla}_{e,\mathbf{F}}\right]_{(i,j)} = \frac{\partial \overline{\mathcal{R}}_e\left(\mathbf{F},\mathbf{T}\right)}{\partial F_{(i,j)}^*}\\
				&= -\tr\left(\mathbf{\Gamma}^{-1}\mathbf{\Gamma}_{(i,j)}\right) + \tr\left[\widetilde{\mathbf{\Phi}}\left(\delta F_{(i,j)}\hat{\delta}(i,j) + \delta_{(i,j)}'\widetilde{\mathbf{D}}\right)\right] \\
				&\quad - \sigma^2 M \left(\delta_{(i,j)}' \tilde{\delta} + \delta \tilde{\delta}_{(i,j)}'\right).
			\end{split}
		\end{equation}
		Here, $F_{(i,j)}$ represents the $(i,j)$th entry of $\mathbf{F}$.
	\end{proposition}

	\textbf{\emph{Proof}}: 	From \eqref{App1}, we can see that $\overline{\mathcal{R}}_{e}$ is a sum of three terms, e.g., $\log \left\vert\sigma^{2}\mathbf{\Gamma}\right\vert$, $\log \left\vert\mathbf{I}+\delta\widetilde{\mathbf{D}}\right\vert$, and $- \sigma^2 M \delta \tilde{\delta}$. Accordingly, the proof is organized into three parts, in each of which we compute the derivative of a single term.

	Before proceeding, we establish several intermediate results.
	Define  $\mathbf{\Xi} = \mathbf{\Lambda}^{\frac{1}{2}}\mathbf{G}_{\mathrm{LoS}}^\HH\mathbf{F}	\widetilde{\mathbf{\Phi}}\mathbf{F}^\HH\mathbf{G}_{\mathrm{LoS}}\mathbf{\Lambda}^{\frac{1}{2}}$. Then, we have
	\begin{equation}
		\begin{split}
			&\mathbf{\Xi}_{(i,j)}'=\frac{\partial \mathbf{\Xi}}{\partial F_{(i,j)}^*} =\mathbf{B}\widetilde{\mathbf{\Phi}}\mathbf{e}_j\mathbf{e}_i^\TT	\mathbf{G}_{\mathrm{LoS}}\mathbf{\Lambda}^{\frac{1}{2}}+\mathbf{B}	\widetilde{\mathbf{\Phi}}_{i,j}'\mathbf{B}^\HH.
		\end{split}
	\end{equation}
	Define $\widetilde{\mathbf{\Xi}} = \mathbf{F}^\HH\mathbf{G}_{\mathrm{LoS}}\mathbf{\Lambda}^{\frac{1}{2}}\mathbf{\Phi}\mathbf{\Lambda}^{\frac{1}{2}}\mathbf{G}_{\mathrm{LoS}}^\HH\mathbf{F}$. Then, we have
	\begin{equation}
		\begin{split}
			&\widetilde{\mathbf{\Xi}}_{(i,j)}'=\frac{\partial  \widetilde{\mathbf{\Xi}}}{\partial F_{(i,j)}^*} =\mathbf{e}_j\mathbf{e}_i^\TT \mathbf{G}_{\mathrm{LoS}}\mathbf{\Lambda}^{\frac{1}{2}}\mathbf{\Phi}  \mathbf{B} +\mathbf{B}^\HH \mathbf{\Phi}_{(i,j)}'\mathbf{B}.
		\end{split}
	\end{equation}
	
	Therefore, by differentiating the equations in \eqref{fpequ_0}, we obtain the following expressions.
	The derivative of $\delta$ w.r.t.  $F_{(i,j)}$ is given by
	\begin{equation}
		\begin{split}
			\delta_{(i,j)}'&= \frac{\partial \delta}{\partial F_{(i,j)}^*}= \frac{1}{M} \tr \left(\mathbf{D}\mathbf{\Gamma}_{(i,j)}'\right).
		\end{split}
	\end{equation}
	The derivative of $\tilde{\delta}$ w.r.t.  $F_{(i,j)}$ is given by
	\begin{equation}
		\begin{split}
			\tilde{\delta}_{(i,j)}'&= \frac{\partial \tilde{\delta}}{\partial F_{(i,j)}^*}=\frac{1}{M} \frac{\partial \tr\left(\widetilde{\mathbf{D}}\widetilde{\mathbf{\Gamma}}\right)}{\partial F_{(i,j)}^*}\\
			&= \frac{1}{M}\tr\left(\widetilde{\mathbf{D}}\widetilde{\mathbf{\Gamma}}_{(i,j)}'\right) + \frac{1}{M} F_{(i,j)} \hat{\delta}(i,j).
		\end{split}
	\end{equation}
	The derivative of $\mathbf{\Gamma}$ w.r.t.  $F_{(i,j)}$ is given by
	\begin{equation}
		\begin{split}
			&\mathbf{\Gamma}_{(i,j)}'= \frac{\partial \mathbf{\Gamma}}{\partial F_{(i,j)}^*}=\frac{\partial \left(\sigma^2 \mathbf{\Phi}^{-1}+\mathbf{\Xi}\right)^{-1}}{\partial F_{(i,j)}^*}\\
			&= - \mathbf{\Gamma}\left(\sigma^2\tilde{\delta}_{(i,j)}' \mathbf{D}+ \mathbf{\Xi}_{(i,j)}' \right)\mathbf{\Gamma}.
		\end{split}
	\end{equation}
	The derivative of $\tilde{\gamma}$ w.r.t.  $F_{(i,j)}$ is given by
	\begin{equation}
		\begin{split}
			&\widetilde{\mathbf{\Gamma}}_{(i,j)}'= \frac{\partial \widetilde{\mathbf{\Gamma}}}{\partial F_{(i,j)}^*} =\frac{\partial \left(\sigma^2 \widetilde{\mathbf{\Phi}}^{-1}+ \widetilde{\mathbf{\Xi}}\right)^{-1}}{\partial F_{(i,j)}^*} \\
			&= - \widetilde{\mathbf{\Gamma}} \left(\sigma^2 \delta F_{(i,j)}\hat{\delta}(i,j) + \sigma^2\delta_{(i,j)}'\widetilde{\mathbf{D}}+ \widetilde{\mathbf{\Xi}}_{(i,j)}' \right)\widetilde{\mathbf{\Gamma}}.
		\end{split}
	\end{equation}
	The deriviate of $\mathbf{\Phi}$ w.r.t. $F_{(i,j)}$ is given by
	\begin{equation}
		\begin{split}
			\mathbf{\Phi}_{(i,j)}' = \frac{\partial \mathbf{\Phi}}{\partial F_{(i,j)}^*} = - \mathbf{\Phi} \left( \tilde{\delta}_{(i,j)}' \mathbf{D} \right) \mathbf{\Phi}.
		\end{split}
	\end{equation}
	The derivative of $\tilde{\phi}$ w.r.t. the $(i,j)$th entry of $\Omega$, denoted by $F_{(i,j)}$, is given by
	\begin{equation}
		\begin{split}
			\widetilde{\mathbf{\Phi}}_{(i,j)}' = \frac{\partial \widetilde{\mathbf{\Phi}}}{\partial F_{(i,j)}^*} = -\widetilde{\mathbf{\Phi}} \left(\delta F_{(i,j)}\hat{\delta}(i,j) + \delta_{(i,j)}'\widetilde{\mathbf{D}}\right)\widetilde{\mathbf{\Phi}}.
		\end{split}
	\end{equation}

	In the following, we give the derivative of each term in \eqref{App1}:
a) First term:
\begin{equation}
	\begin{split}
		&\frac{\partial \log|\sigma^2\mathbf{\Gamma}|}{\partial F_{(i,j)}^*} = \tr\left(\mathbf{\Gamma}^{-1}\mathbf{\Gamma}_{(i,j)}\right),
	\end{split}
\end{equation}
b) Second term:
\begin{equation}
\begin{split}
	\frac{\partial \log \left\vert\mathbf{I}+\delta\widetilde{\mathbf{D}}\right\vert}{\partial F_{(i,j)}^*} = \tr\left[\widetilde{\mathbf{\Phi}}\left(\delta F_{(i,j)}\hat{\delta}(i,j) + \delta_{(i,j)}'\widetilde{\mathbf{D}}\right) \right],
\end{split}
\end{equation}
c) Third term:
\begin{equation}
	\begin{split}
		&\frac{\partial \sigma^2 M \delta \tilde{\delta}}{\partial F_{(i,j)}^*} = \sigma^2 M \left(\delta_{(i,j)}' \tilde{\delta} + \delta \tilde{\delta}_{(i,j)}'\right),
	\end{split}
\end{equation}
which completes the proof \hfill $\blacksquare$.

	Given $-\overline{\mathcal{R}}_{e}$ is concave, it can be lower-bounded by its first-order Taylor expansion around the results at the previous iteration $\mathbf{F}_{k,t}$, i.e., 
	\begin{equation}
		\begin{split}
			-\overline{\mathcal{R}}_{e}\left(\mathbf{F};\mathbf{T}_k\right) \geq -\overline{\mathcal{R}}_{e}\left(\mathbf{F}_{k,t};\mathbf{T}_k\right) - \bm{\nabla}_{e,\mathbf{F}_{k,t}}^\HH \left(\mathbf{F} - \mathbf{F}_{k,t}\right),
		\end{split}
	\end{equation}
	where $\bm{\nabla}_{e,\mathbf{F}_{k,t}} = \left.\bm{\nabla}_{e,\mathbf{F}} \right\vert_{\mathbf{F} = \mathbf{F}_{k,t}}$.

Thus, the surrogate function is constructed as
\begin{equation}
	\mathcal{G}\left(\mathbf{F}\right) = \mathcal{R}_{b}\left(\mathbf{F};\mathbf{T}_k\right)-\overline{\mathcal{R}}_{e}\left(\mathbf{F}_{k,t};\mathbf{T}_k\right) - \bm{\nabla}_{e,\mathbf{F}_{k,t}}^\HH \left(\mathbf{F} - \mathbf{F}_{k,t}\right).
\end{equation}
Its gradient at the point $\mathbf{F} = \mathbf{F}_{k,t}$ is given by 
\begin{equation}
	\begin{split}
	&\bm{\nabla}_{\mathbf{F}_{k,t}} =\left.\frac{\partial \mathcal{G}\left(\mathbf{F}\right)}{\partial \mathbf{F}}\right\vert_{\mathbf{F} = \mathbf{F}_{k,t}} \\
	&= \sigma^{-2} \mathbf{H}\left(\mathbf{I}+\sigma^{-2}\mathbf{H}^\HH \mathbf{F}_{k,t}\mathbf{F}_{k,t}^\HH\mathbf{H}\right)^{-1}\mathbf{H}^\HH \mathbf{F}_{k,t} - \bm{\nabla}_{e,\mathbf{\Omega}_{k,t}}.
	\end{split}
\end{equation}

	Then, we update $\mathbf{F}$ by a gradient step, i.e.,
	\begin{equation}\label{update_Omega}
		\widetilde{\mathbf{F}}_{k,t+1} = \mathbf{F}_{k,t} - \alpha_{k,t} \bm{\nabla}_{\mathbf{F}_{k},t}, 
	\end{equation}
	where 
	the step size $\alpha_{k,t}$ is determined by an Armijo line search. Finally, a retraction maps the provisional iterate back to the feasible set. Here, we enforce the trace constraint by
	\begin{equation}\label{retraction1}
		\begin{split}
			\mathbf{F}_{k,t+1} = \left\{
			\begin{matrix}
				\widetilde{\mathbf{F}}_{k,t+1}, & \mathrm{if}\;  \left\Vert\widetilde{\mathbf{F}}_{k,t+1}\right\Vert^2 \leq P \\
				\frac{\sqrt{P} }{\left\Vert\widetilde{\mathbf{F}}_{k,t+1}\right\Vert}\widetilde{\mathbf{F}}_{k,t+1}, &\mathrm{otherwise}
			\end{matrix}
			\right. .
		\end{split}
	\end{equation}

	The proposed algorithm is summarized in \textbf{\emph{Algorithm \ref{alg2}}}.
	The convergence of the proposed method is guaranteed by \cite[Theorem 4.3.1]{absil2008optimization}

	\begin{algorithm}[!t]
		\caption{The proposed GP-based method for solving (\ref{prob_221})}
		\begin{algorithmic}[1]
			\STATE Initialize $\mathbf{F}_{k,0} =\mathbf{F}_{k} $.
			\STATE \textbf{Repeat}
			\STATE \hspace{0.5cm}Compute the gradient $\bm{\nabla}_{\mathbf{F}}$ via \eqref{Delta_Omega}. 
			\STATE \hspace{0.5cm}Compute $\alpha_{k,t}$ via the Armijo line search step.
			\STATE \hspace{0.5cm}Update $\widetilde{\mathbf{F}}_{k,t+1}$ via \eqref{update_Omega}.
			\STATE \hspace{0.5cm}Update $\mathbf{F}_{k,t+1}$ via the retraction defined in \eqref{retraction1}.
			\STATE \hspace{0.5cm} $t \gets t+1$.
			\STATE \textbf{Until} Convergence criterion is met. 
			\STATE \textbf{return} $\mathbf{F}_{k+1}=\mathbf{F}_{k,t}$.
		\end{algorithmic}
		\label{alg2}
	\end{algorithm}

	\subsection{Update $\mathbf{T}$ with Given $\mathbf{F}_{k+1}$}
	In this subproblem, we aim to optimize $\mathbf{t}_n$ sequentially with given $\mathbf{F}_{k+1}$. 
	Unlike the previous problem, the two parts of the objective function, i.e., $\mathcal{R}_b(\mathbf{t}_n)$ and $- \overline{\mathcal{R}}_e(\mathbf{t}_n)$, are non-convex w.r.t. the variable $\mathbf{t}_n$. One effective strategy for solving such non-convex problems is the MM framework, which iteratively minimizes a surrogate function that majorizes the original objective. Nevertheless, the construction of an appropriate surrogate function typically necessitates the Hessian matrix of $ \overline{\mathcal{R}}_e(\mathbf{t}_n)$. For problem \eqref{prob_222}, obtaining the Hessian is extremely challenging, as its objective function involves solving a fixed-point equations, leading to expressions that are both computationally prohibitive and analytically intractable. 
	
	To address this difficulty, we propose to leverage the AMSGrad optimization algorithm to update $\mathbf{t}_n$. As a sophisticated first-order gradient-based method, AMSGrad combines the advantages of both root mean square propagation (RMSProp) and momentum-based stochastic gradient descent. Its adoption in our work is motivated by its superior performance in non-convex optimization landscapes, specifically: 
	1) it only requires the first-order gradient; 
	2) its minimal computational overhead and memory-efficient profile, making it suitable for real-time applications; 
	3) its proven scalability in high-dimensional parameter spaces, which aligns with the complexity of our optimization problem; and 
	4) its ability to dynamically adjust learning rates, which ensures stable convergence even when the objective function exhibits sparse or non-stationary gradients \cite{reddi2019convergence}.
	
	Next, we present the following proposition, which provide the gradient of $\overline{\mathcal{R}}_e\left(\mathbf{F},\mathbf{T}\right)$ w.r.t. $\mathbf{t}_{n}$.

	\begin{proposition}\label{grad_tn}
		By differentiating the equations in \eqref{fpequ_0} w.r.t. the $i$th entry of $\mathbf{t}_n$, denoted by $t_{n,i}$, we establish the following fixed-point equations.
		\begin{equation}\label{fpequ_grad_2}
			\begin{split}
				\left\{
				\begin{array}{l}
					\delta_{(n,i)}'= \frac{1}{M} \tr \left(\mathbf{D}\mathbf{\Gamma}_{(n,i)}'\right)\\
					\tilde{\delta}_{(n,i)}'=\frac{1}{M} \tr \left(\widetilde{\mathbf{D}}\widetilde{\mathbf{\Gamma}}_{(n,i)}'\right)\\
				\end{array}\right.,
			\end{split}
		\end{equation}
		where
		\begin{subequations}
			\begin{equation}
				\begin{split}
					&\mathbf{\Gamma}_{(n,i)}'=  - \mathbf{\Gamma}\left(\sigma^2\tilde{\delta}_{(n,i)}' \mathbf{D}+ \mathbf{\Xi}_{(n,i)}' \right)\mathbf{\Gamma},
				\end{split}
			\end{equation}
			\begin{equation}
				\begin{split}
					\widetilde{\mathbf{\Gamma}}_{(n,i)}' &=  - \widetilde{\mathbf{\Gamma}} \left( \sigma^2\delta_{(n,i)}' \widetilde{\mathbf{D}} + \widetilde{\mathbf{\Xi}}_{(n,i)}' \right)\widetilde{\mathbf{\Gamma}},
				\end{split}
			\end{equation}
			\begin{equation}
				\begin{split}
					\mathbf{\Phi}_{(n,i)}' = - \mathbf{\Phi} \left( \tilde{\delta}_{(n,i)}' \mathbf{D} \right) \mathbf{\Phi},
				\end{split}
			\end{equation}
			\begin{equation}
				\begin{split}
					\widetilde{\mathbf{\Phi}}_{(n,i)}' = -  \widetilde{\mathbf{\Phi}}\left(\delta_{(n,i)}'\widetilde{\mathbf{D}}\right)\widetilde{\mathbf{\Phi}},
				\end{split}
			\end{equation}
			\begin{equation}
				\begin{split}
					&\mathbf{\Xi}_{(n,i)}'
					=\mathbf{B}	\widetilde{\mathbf{\Phi}}_{(n,i)}'\mathbf{B}^\HH +2\Re\left(\mathbf{\Lambda}^{\frac{1}{2}}\tilde{\mathbf{g}}_{n,i}'\mathbf{e}_n^\TT\mathbf{F}	\widetilde{\mathbf{\Phi}}\mathbf{B}^\HH \right),
				\end{split}
			\end{equation}
			\begin{equation}
				\begin{split}
					&\widetilde{\mathbf{\Xi}}_{(n,i)}'=\mathbf{B}^\HH\mathbf{\Phi}_{(n,i)}' \mathbf{B} +2\Re\left(\mathbf{B}^\HH\mathbf{\Phi}\mathbf{\Lambda}^{\frac{1}{2}}\tilde{\mathbf{g}}_{n,i}'\mathbf{e}_n^\TT \mathbf{F}\right) .
				\end{split}
			\end{equation}
		\end{subequations}
		The gradient of $\overline{\mathcal{R}}_e\left(\mathbf{F},\mathbf{T}\right)$ w.r.t. $\mathbf{t}_{n}$ is denoted by $\bm{\nabla}_{e,\mathbf{t}_{n}} \triangleq \frac{\partial \overline{\mathcal{R}}\left(\mathbf{F},\mathbf{T}\right)}{\partial \mathbf{t}_{n}}$, whose $i$th entry is given by
		\begin{equation}
			\begin{split}
				[\bm{\nabla}_{e,\mathbf{t}_{n}}]_{i}&= -\tr\left(\mathbf{\Gamma}^{-1}\mathbf{\Gamma}_{(n,i)}\right) + \tilde{\phi}\delta_{(n,i)}' \tr\left(\mathbf{F}\mathbf{F}^\HH\right) \\
				& \quad- \sigma^2   M  \left(\delta_{(n,i)}' \tilde{\delta} + \delta \tilde{\delta}_{(n,i)}'\right).
			\end{split}
		\end{equation}
		Here, $t_{n,i}$ represents the $i$th entry of $\mathbf{t}_{n}$.
	\end{proposition}

	\textbf{\emph{Proof}}: 	Similarly, we begin with establishing several intermediate results.
	For simplicity, we define 
	\begin{equation}
		\widetilde{\mathbf{G}} = \mathbf{G}_{\mathrm{LoS}}^\HH = \left[\tilde{\mathbf{g}}(\mathbf{t}_1),\tilde{\mathbf{g}}(\mathbf{t}_2),\cdots,\tilde{\mathbf{g}}(\mathbf{t}_N)\right],
	\end{equation}
	where
	\begin{equation}
		\tilde{\mathbf{g}}(\mathbf{t}_n) = \left[e^{-j\frac{2\pi}{\lambda}\mathbf{t}_n^\TT\bm{\rho}_{e,1}},e^{-j\frac{2\pi}{\lambda}\mathbf{t}_n^\TT\bm{\rho}_{e,2}},\cdots,e^{-j\frac{2\pi}{\lambda}\mathbf{t}_n^\TT\bm{\rho}_{e,M}}\right]^\TT.
	\end{equation}
	Then, we have
	\begin{equation}
		\widetilde{\mathbf{G}}_{(n,i)}' \triangleq \frac{\partial \widetilde{\mathbf{G}}}{\partial t_{n,i}} = \left[\mathbf{0},\mathbf{0},\cdots,\tilde{\mathbf{g}}'(\mathbf{t}_n),\cdots,\mathbf{0}\right]=\tilde{\mathbf{g}}_{n,i}'\mathbf{e}_n^\TT,
	\end{equation}
	where $\tilde{\mathbf{g}}_{(n,i)}'\triangleq \frac{\partial \tilde{\mathbf{g}}(\mathbf{t}_n)}{\partial t_{n,i}}$ is defined by the gradient of $\tilde{\mathbf{g}}(\mathbf{t}_n)$ w.r.t. $t_{n,i}$, i.e.,
	\begin{equation}\nonumber
		\tilde{\mathbf{g}}_{(n,i)}' =-j\frac{2\pi}{\lambda}\left[[\bm{\rho}_{e,1}]_i e^{-j\frac{2\pi}{\lambda}\mathbf{t}_n^\TT\bm{\rho}_{e,1}},\cdots,[\bm{\rho}_{e,M}]_i e^{-j\frac{2\pi}{\lambda}\mathbf{t}_n^\TT\bm{\rho}_{e,M}}\right]^\TT.
	\end{equation}

	Then, we have
	\begin{equation}
		\begin{split}
			&\mathbf{\Xi}_{(n,i)}'=\frac{\partial \mathbf{\Xi}}{\partial t_{n,i}} = \frac{\partial \mathbf{\Lambda}^{\frac{1}{2}}\widetilde{\mathbf{G}}\mathbf{F}	\widetilde{\mathbf{\Phi}}\mathbf{F}^\HH \widetilde{\mathbf{G}}^\HH\mathbf{\Lambda}^{\frac{1}{2}}}{\partial t_{n,i}}\\
			&=\mathbf{\Lambda}^{\frac{1}{2}}\widetilde{\mathbf{G}}\mathbf{F}	\widetilde{\mathbf{\Phi}}_{(n,i)}'\mathbf{F}^\HH \widetilde{\mathbf{G}}^\HH\mathbf{\Lambda}^{\frac{1}{2}}\\
			&+\mathbf{\Lambda}^{\frac{1}{2}}\widetilde{\mathbf{G}}_{(n,i)}'\mathbf{F}	\widetilde{\mathbf{\Phi}}\mathbf{F}^\HH\widetilde{\mathbf{G}}^\HH\mathbf{\Lambda}^{\frac{1}{2}} + \mathbf{\Lambda}^{\frac{1}{2}}\widetilde{\mathbf{G}}\mathbf{F}	\widetilde{\mathbf{\Phi}}\mathbf{F}^\HH\left(\widetilde{\mathbf{G}}_{(n,i)}'\right)^\HH\mathbf{\Lambda}^{\frac{1}{2}},
		\end{split}
	\end{equation}
	\begin{equation}
		\begin{split}
			&\widetilde{\mathbf{\Xi}}_{(n,i)}'=\frac{\partial  \widetilde{\mathbf{\Xi}}}{\partial t_{n,i}} = \frac{\partial\mathbf{F}^\HH\widetilde{\mathbf{G}}^\HH\mathbf{\Lambda}^{\frac{1}{2}}\mathbf{\Phi}\mathbf{\Lambda}^{\frac{1}{2}}\widetilde{\mathbf{G}} \mathbf{F}}{\partial t_{n,i}}\\
			&=\mathbf{F}^\HH\widetilde{\mathbf{G}}^\HH\mathbf{\Lambda}^{\frac{1}{2}}\mathbf{\Phi}_{(n,i)}'\mathbf{\Lambda}^{\frac{1}{2}}\widetilde{\mathbf{G}} \mathbf{F}  \\ 
			&+\mathbf{F}^\HH\widetilde{\mathbf{G}}^\HH\mathbf{\Lambda}^{\frac{1}{2}}\mathbf{\Phi}\mathbf{\Lambda}^{\frac{1}{2}}\widetilde{\mathbf{G}}_{(n,i)}' \mathbf{F} + \mathbf{F}^\HH\left(\widetilde{\mathbf{G}}_{(n,i)}'\right)^\HH\mathbf{\Lambda}^{\frac{1}{2}}\mathbf{\Phi}\mathbf{\Lambda}^{\frac{1}{2}}\widetilde{\mathbf{G}} \mathbf{F}.
		\end{split}
	\end{equation}
	Then, the derivative of $\delta$, $\tilde{\delta}$, $\mathbf{\Gamma}$, $\widetilde{\mathbf{\Gamma}}$, $\mathbf{\Phi}$, and $\widetilde{\mathbf{\Phi}}$ w.r.t.  $t_{n,i}$ are respectively given by
	\begin{equation}
		\begin{split}
			\delta_{(n,i)}'&= \frac{\partial \delta}{\partial t_{n,i}}= 	\frac{1}{M} \tr \left(\mathbf{D}\mathbf{\Gamma}_{(n,i)}'\right),
		\end{split}
	\end{equation}
	\begin{equation}
		\begin{split}
			\tilde{\delta}_{(n,i)}'&= \frac{\partial \tilde{\delta}}{\partial t_{n,i}} =  \frac{1}{M} \tr \left(\widetilde{\mathbf{D}}\widetilde{\mathbf{\Gamma}}_{(n,i)}'\right),
		\end{split}
	\end{equation}
	\begin{equation}
		\begin{split}
			&\mathbf{\Gamma}_{(n,i)}'= \frac{\partial \mathbf{\Gamma}}{\partial t_{n,i}}= - \mathbf{\Gamma}\left(\sigma^2\tilde{\delta}_{(n,i)}' \mathbf{D}+ \mathbf{\Xi}_{(n,i)}' \right)\mathbf{\Gamma},
		\end{split}
	\end{equation}
	\begin{equation}
		\begin{split}
			\widetilde{\mathbf{\Gamma}}_{(n,i)}' &= \frac{\partial \widetilde{\mathbf{\Gamma}}}{\partial t_{n,i}} = - \widetilde{\mathbf{\Gamma}} \left( \sigma^2\delta_{(n,i)}' \widetilde{\mathbf{D}} + \widetilde{\mathbf{\Xi}}_{(n,i)}' \right)\widetilde{\mathbf{\Gamma}},
		\end{split}
	\end{equation}
	\begin{equation}
		\begin{split}
			\mathbf{\Phi}_{(n,i)}' = \frac{\partial \mathbf{\Phi}}{\partial t_{n,i}} = - \mathbf{\Phi} \left( \tilde{\delta}_{(n,i)}' \mathbf{D} \right) \mathbf{\Phi},
		\end{split}
	\end{equation}
	\begin{equation}
		\begin{split}
			\widetilde{\mathbf{\Phi}}_{(n,i)}' = \frac{\partial \widetilde{\mathbf{\Phi}}}{\partial t_{n,i}} = -  \widetilde{\mathbf{\Phi}}\left(\delta_{(n,i)}'\widetilde{\mathbf{D}}\right)\widetilde{\mathbf{\Phi}}.
		\end{split}
	\end{equation}

	In the following, we give the derivative of each term in \eqref{App1}:
	a) First term:
	\begin{equation}
		\begin{split}
			&\frac{\partial \log|\sigma^2\mathbf{\Gamma}|}{\partial t_{n,i}} = \tr\left(\mathbf{\Gamma}^{-1}\mathbf{\Gamma}_{(n,i)}\right),
		\end{split}
	\end{equation}
	b) Second term:
	\begin{equation}
		\begin{split}
			\frac{\partial \log \left\vert \mathbf{I} + \delta \widetilde{\mathbf{D}}\right\vert}{\partial t_{n,i}} =\delta_{(n,i)}' \tr\left(\widetilde{\mathbf{\Phi}} \widetilde{\mathbf{D}}\right),
		\end{split}
	\end{equation}
	c) Third term:
		\begin{equation}
		\begin{split}
			\frac{\partial \sigma^2 M \delta \tilde{\delta}}{\partial t_{n,i}} = \sigma^2 M \left(\delta_{(n,i)}' \tilde{\delta} + \delta \tilde{\delta}_{(n,i)}'\right),
		\end{split}
	\end{equation}
	which completes the proof. \hfill $\blacksquare$
	
Given the gradient of $\overline{\mathcal{R}}_e(\mathbf{t}_n)$ provided in \textbf{\emph{Proposition \ref{grad_tn}}}, we next derive the gradient of $\mathcal{R}_b(\mathbf{t}_n)$. To begin with, we introduce the steering-like vector $\mathbf{a}(\mathbf{t}_n)$ defined as
	\begin{equation}
		\mathbf{a}(\mathbf{t}_n) = \left[ e^{j\frac{2\pi}{\lambda}\mathbf{t}_n^{\mathrm{T}}\bm{\rho}_{b,1}}, \dots, e^{j\frac{2\pi}{\lambda}\mathbf{t}_n^{\mathrm{T}}\bm{\rho}_{b,L}} \right]^{\mathrm{T}} \in \mathbb{C}^{L\times 1}.
	\end{equation}
	Then, the $l$th entry of $\mathbf{a}_{n,i}' = \frac{\partial \mathbf{a}(\mathbf{t}_n)}{\partial t_{n,i}}$ and $\mathbf{a}_{n,(i,j)}'' = \frac{\partial^2 \mathbf{a}(\mathbf{t}_n)}{\partial t_{n,i} \partial t_{n,j}}$ is respectively given by
    \begin{subequations}
	\begin{equation}\label{anil_diff}
	\left[\mathbf{a}_{n,i}'\right]_l = j\frac{2\pi}{\lambda}[\bm{\rho}_{b,l}]_ie^{j\frac{2\pi}{\lambda}\mathbf{t}_n^\TT\bm{\rho}_{b,l}}.
	\end{equation}
	\begin{equation}
		\left[\mathbf{a}_{n,(i,j)}''\right]_l = -\frac{4\pi^2}{\lambda^2}[\bm{\rho}_{b,l}]_i [\bm{\rho}_{b,l}]_j e^{j\frac{2\pi}{\lambda}\mathbf{t}_n^\TT\bm{\rho}_{b,l}}.
	\end{equation}
    \end{subequations}
	
	Meanwhile, the gradient of $\mathcal{R}_{b}\left(\mathbf
	{t}_n\right)$ w.r.t. $\mathbf{t}_{n}$ is denoted by $\bm{\nabla}_{b,\mathbf{t}_{n}}$, whose $i$th entry is given by
	\begin{equation}\label{grad_tn_def}
		\begin{split}
			&\left[\bm{\nabla}_{b,\mathbf{t}_{n}}\right]_{i} =  \frac{\partial \mathcal{R}_{b}\left(\mathbf{t}_n\right)}{\partial t_{n,i}}\\
			&= \frac{2}{\sigma^{2}} \tr\left[\left(\mathbf{I}+\sigma^{-2} \mathbf{H}^\HH \mathbf{F}_{k}\mathbf{F}_{k}^\HH  \mathbf{H} \right)^{-1}\Re\left(\mathbf{H}^\HH \mathbf{F}_{k}\mathbf{F}_{k}^\HH \mathbf{H}_{n,i}'  \right)\right],
		\end{split}
	\end{equation}
	where
	\begin{equation}
		\mathbf{H}_{n,i}' =  \mathbf{e}_n \left(\mathbf{a}_{n,i}'\right)^\TT \mathbf{\Lambda}_b \mathbf{A}_R^\HH.
	\end{equation}

	Note that problem \eqref{prob_222} is equivalent to
	\begin{equation}\label{prob_2222}
		\begin{split}
				\mathbf{t}_{n,k+1} &= \arg \min_{\mathbf{t}_{n}\in \mathcal{S}_{T}} \mathcal{L}(\mathbf{t}_n),
			\end{split}
	\end{equation}
	where 
	\begin{equation}
		\mathcal{L}(\mathbf{t}_n) = -\mathcal{R}_{b} \left(\mathbf{t}_n;\mathbf{F}_{k+1},\mathbf{T}_{k}\right) + \overline{\mathcal{R}}_{e} \left(\mathbf{t}_n;\mathbf{F}_{k+1},\mathbf{T}_{k}\right).
	\end{equation}
	Then, the overall gradient of the objective function in \eqref{prob_2222} is given by
\begin{equation}\label{grad_tn_minor}
\bm{\nabla}_{\mathbf{t}_{n}}= - \bm{\nabla}_{b,\mathbf{t}_{n}} + \bm{\nabla}_{e,\mathbf{t}_{n}}.
\end{equation}
	
	Following \cite{reddi2019convergence}, the first-order and second-order moments are given by
\begin{equation}\label{hat_m_t}
		\widehat{\mathbf{m}}_t = \beta_{1,t} \widehat{\mathbf{m}}_{t-1} + (1-\beta_{1,t}) \widehat{\mathbf{g}}_{t},
\end{equation}
\begin{equation}\label{hat_v_t}
	\widehat{\mathbf{v}}_t = \beta_{2} \widehat{\mathbf{v}}_{t-1} + (1-\beta_{2}) \left\vert\widehat{\mathbf{g}}_{t}\right\vert^2,
\end{equation}
	where $\widehat{\mathbf{g}}_{t} =  \bm{\nabla}_{\mathbf{t}_{n}}|_{\mathbf{t}_{n} = \mathbf{t}_{n,(k,t)}}$ denotes the gradient at the current iteration, and $\beta_2 \in (0,1)$. 
	The intermediate parameter is set as 
	\begin{equation}
		\beta_{1,t} = \beta_1 \lambda_1^{t-1},
	\end{equation}
	where $\beta_1, \lambda_1 \in (0,1)$.
	
	Then, the variable $\mathbf{t}_{n,(k,t+1)}$ can be updated by 
	\begin{equation}
		\mathbf{t}_{n,(k,t+1)} = \min_{\mathbf{t}_{n}\in \mathcal{S}_{T}} \left\Vert \widehat{\mathbf{V}}_t^{\frac{1}{2}}\left(\mathbf{t}_n - \left(\mathbf{t}_{n,(k,t)} - \alpha_t \widehat{\mathbf{V}}_t^{-\frac{1}{2}} \widehat{\mathbf{m}}_t \right)\right) \right\Vert^2,
	\end{equation}
	where $\widehat{\mathbf{V}}_t = \diag\left(	\widehat{\mathbf{v}}_t\right)$, and the step size $\alpha_t$ is typically set to $\alpha/\sqrt{t}$ for some constant $\alpha$. This problem is equivalent to a quadratic programming (QP) problem, i.e.,
	\begin{equation}\label{prob_2223}
		\mathbf{t}_{n,(k,t+1)} = \min_{\mathbf{t}_{n}\in \mathcal{S}_{T}} \mathbf{t}_n^\TT \widehat{\mathbf{V}}_t \mathbf{t}_n + 2 \widehat{\mathbf{f}}_t^\TT \mathbf{t}_n
	\end{equation}
	where  
\begin{equation}\label{f_2}
	\widehat{\mathbf{f}}_t = - \widehat{\mathbf{V}}_t \mathbf{t}_{n,(k,t)} + \alpha_t \widehat{\mathbf{V}}_t^{\frac{1}{2}} \widehat{\mathbf{m}}_t.
\end{equation} 
	This problem \eqref{prob_2223} can be efficiently solved by using quadprog \cite{cvx}. The proposed AMSGrad method is summarised in \textbf{Algorithm \ref{alg3}}.

	By far, we have developed an AMSGrad-based optimization scheme to update the antenna positions $\mathbf{t}_n$. However, it is worth noting that the constructed surrogate function may not strictly adhere to the fundamental properties required by the MM framework, such as the tangency and majorization conditions. Consequently, the convergence guarantee inherently associated with standard MM algorithms becomes questionable in this context. To provide a rigorous theoretical foundation for the proposed method, we proceed to analyze the convergence behavior of the AMSGrad-based algorithm in the following proposition.

	To facilitate the convergence analysis, we define the following regret function for any generated sequence $\{\mathbf{t}_{n,(k,t)}\}_{t=1}^T$ as
	\begin{equation}
		R_T = \sum_{t=1}^{T} \left[ \mathcal{L}(\mathbf{t}_{n,(k,t)}) - \mathcal{L}(\mathbf{t}_{n,(k,\star)}) \right],
	\end{equation}
	where $\mathbf{t}_{n,(k,\star)} \in \mathcal{S}_T$ denotes the optimal stationary point within the feasible set $\mathcal{S}_T$. Essentially, the regret $R_T$ quantifies the cumulative performance gap between the objective values evaluated at the $t$-th iteration and the optimal benchmark over $T$ iterations.

	\begin{proposition}\label{ConverAMSG}
		The regret is bounded by
\begin{equation}\label{regret01}
	\begin{split}
		R_T \leq &\frac{2 D_\infty^2 G_\infty \sqrt{T}}{\alpha \left(1 - \beta_1\right)} + \frac{\beta_1 D_\infty^2 G_\infty }{(1-\beta_1)^2 \left(1-\lambda_1\right)^2}\\ &+\frac{2\alpha G_\infty \sqrt{1+\log T}\sqrt{T}}{\left(1-\beta_1\right)^2\left(1-\gamma\right)\sqrt{1-\beta_2}},
	\end{split}
\end{equation}
where %$D_{\infty} = 2\sqrt{D_x^2 + D_y^2},$ and $ G_\infty = \frac{8\pi \sqrt{M}}{\lambda} $. 
\begin{equation}
\begin{aligned}
	&D_{\infty} = 2\sqrt{D_x^2 + D_y^2}, \\ &G_\infty = \frac{8\pi \sqrt{M}}{\lambda}.
\end{aligned}
\end{equation}
	\end{proposition}
	
	\textbf{\emph{Proof}}:  The regret of the AMSGrad is given by \cite[Corollary 1]{reddi2019convergence} 
	\begin{equation}\label{regret00}
		\begin{split}
			R_T \leq &\frac{D_\infty^2 \sqrt{T}}{\alpha \left(1 - \beta_1\right)} \sum_{i=1}^{2} \left[\widehat{\mathbf{v}}_T\right]_i^{\frac{1}{2}} + \frac{\beta_1 D_\infty^2 G_\infty }{(1-\beta_1)^2 \left(1-\lambda\right)^2}\\ &+\frac{\alpha \sqrt{1+\log T}}{\left(1-\beta_1\right)^2\left(1-\gamma\right)\sqrt{1-\beta_2}} \sum_{i=1}^{2} \sqrt{\sum_{t=1}^{T}\left[\widehat{\mathbf{g}}_t\right]_i^2} .
		\end{split}
	\end{equation}
	The primary objective of this proof is to characterize $R_T$ within the context of the formulated problem.
	
	In this paper, a rectangular movement constraint is considered for the antennas, where the deployment area is defined by dimensions $D_x \times D_y$. The diameter of the corresponding feasible set $\mathcal{S}_T$, which represents the maximum distance between any two points in the region, is given by $D_{\infty} = 2\sqrt{D_x^2 + D_y^2}$.

	%\subsection{$G_\infty$}
	From \eqref{anil_diff}, we have
	\begin{equation}
		\left[\mathbf{a}_{n,i}'\right]_l = j\frac{2\pi}{\lambda}[\bm{\rho}_{b,l}]_ie^{j\frac{2\pi}{\lambda}\mathbf{t}_n^\TT\bm{\rho}_{b,l}} = j\frac{2\pi}{\lambda}[\bm{\rho}_{b,l}]_i [\mathbf{a}(\mathbf{t}_n)]_l.
	\end{equation}
	By definition of $\bm{\rho}_{b,l}$, we have $\left\vert[\bm{\rho}_{b,l}]_i \right\vert^2 \leq 1.$
	Moreover, we have
	\begin{equation}
		\begin{split}
			\delta_H &\triangleq \frac{1}{\sigma^{2}}\tr\left[\left(\mathbf{I}+\sigma^{-2} \mathbf{H}^\HH \mathbf{F}_{k}\mathbf{F}_{k}^\HH  \mathbf{H} \right)^{-1}\left(\mathbf{H}^\HH \mathbf{F}_{k}\mathbf{F}_{k}^\HH \mathbf{H}  \right)\right] \\
			& \leq \tr \left[\mathbf{I} - \left(\mathbf{I}+\sigma^{-2} \mathbf{H}^\HH \mathbf{F}_{k}\mathbf{F}_{k}^\HH  \mathbf{H} \right)^{-1} \right] \leq M.
		\end{split}
	\end{equation}
	From \eqref{grad_tn_def}, we can see that
	\begin{equation}\label{grad_tn_def2}
		\begin{split}
			&\left\vert\left[\bm{\nabla}_{b,\mathbf{t}_{n}}\right]_{i}\right\vert^2 \\
			&=\left\vert \frac{2}{\sigma^{2}} \tr\left[\left(\mathbf{I}+\sigma^{-2} \mathbf{H}^\HH \mathbf{F}_{k}\mathbf{F}_{k}^\HH  \mathbf{H} \right)^{-1}\Re\left(\mathbf{H}^\HH \mathbf{F}_{k}\mathbf{F}_{k}^\HH \mathbf{H}_{n,i}'  \right)\right]\right\vert^2\\
			& \leq \frac{16\pi^2}{\lambda^2}\max_l\left([\bm{\rho}_{b,l}]_i\right)^2 \cdot\delta_H \leq \frac{16\pi^2M}{\lambda^2}.
		\end{split}
	\end{equation}
	Similarly, we have
	\begin{equation}\nonumber
		\begin{split}
			&\left\vert\left[\bm{\nabla}_{e,\mathbf{t}_{n}}\right]_{i}\right\vert^2 \\
			&=\left\vert \frac{2}{\sigma^{2}} \mathbb{E}\left(\tr\left[\left(\mathbf{I}+\sigma^{-2} \mathbf{G}^\HH \mathbf{F}_{k}\mathbf{F}_{k}^\HH  \mathbf{G} \right)^{-1}\Re\left(\mathbf{G}^\HH \mathbf{F}_{k}\mathbf{F}_{k}^\HH \mathbf{G}_{n,i}'  \right)\right]\right)\right\vert^2\\
			&\overset{(a)}{\leq}\mathbb{E}\left(\left\vert \frac{2}{\sigma^{2}} \tr\left[\left(\mathbf{I}+\sigma^{-2} \mathbf{G}^\HH \mathbf{F}_{k}\mathbf{F}_{k}^\HH  \mathbf{G} \right)^{-1}\Re\left(\mathbf{G}^\HH \mathbf{F}_{k}\mathbf{F}_{k}^\HH \mathbf{G}_{n,i}'  \right]\right)\right\vert^2\right)\\
			& \leq \frac{16\pi^2 M}{\lambda^2}\max_m\left([\bm{\rho}_{e,m}]_i\right)^2  \leq \frac{16\pi^2M}{\lambda^2},
		\end{split}
	\end{equation}
	where $(a)$ comes from the Jensen's inequality, i.e., $|\mathbb{E}(x)|^2 \leq \mathbb{E}(|x^2|)$. 
	Therefore, we have 
	\begin{equation}
		\begin{split}
			&\left\Vert\widehat{\mathbf{g}}_t\right\Vert_\infty = \max_i \left\vert\left[-\bm{\nabla}_{b,\mathbf{t}_{n}}+\bm{\nabla}_{e,\mathbf{t}_{n}}\right]_{i}\right\vert \\
			& \leq \max_i \left\vert\left[\bm{\nabla}_{b,\mathbf{t}_{n}}\right]_{i}\right\vert+\max_i\left\vert\left[\bm{\nabla}_{e,\mathbf{t}_{n}}\right]_{i}\right\vert\leq \frac{8\pi \sqrt{M}}{\lambda} = G_\infty.
		\end{split}
	\end{equation}

	By definition of $\widehat{\mathbf{v}}_t$ in \eqref{hat_v_t}, we have
	\begin{equation}
		\begin{split}
			\widehat{\mathbf{v}}_T &= \beta_{2} \widehat{\mathbf{v}}_{T-1} + \left(1-\beta_{2}\right) \left\vert\widehat{\mathbf{g}}_{T}\right\vert^2,\\
			& = \beta_{2}^2 \widehat{\mathbf{v}}_{T-2} + \beta_{2}\left(1-\beta_{2}\right) \left\vert\widehat{\mathbf{g}}_{T-1}\right\vert^2 + \left(1-\beta_{2}\right) \left\vert\widehat{\mathbf{g}}_{T}\right\vert^2 \\
			& = \cdots= \beta_{2}^{T} \widehat{\mathbf{v}}_{0} + \sum_{t=1}^T\beta_{2}^{T-t}\left(1-\beta_{2}\right) \left\vert\widehat{\mathbf{g}}_{T-t}\right\vert^2 \\
			& \overset{(a)}{=} \left(1-\beta_{2}^T\right) \left\vert\widehat{\mathbf{g}}_{t-1}\right\vert^2,
		\end{split}
	\end{equation}
	where $(a)$ follows from the fact that $\widehat{\mathbf{v}}_{0} = \mathbf{0}$. 
	Then, we have
	\begin{equation}\label{vt12}
		\begin{split}
			\sum_{i=1}^{2} \left[\widehat{\mathbf{v}}_T\right]_i^{\frac{1}{2}} \leq 2\left\Vert\widehat{\mathbf{v}}_t\right\Vert_\infty^{\frac{1}{2}} \leq 2\sqrt{\left(1-\beta_{2}^T\right)} G_\infty \leq 2 G_\infty,
		\end{split}
	\end{equation}
	\begin{equation}\label{gt12}
		\begin{split}
			\sum_{i=1}^{2} \sqrt{\sum_{t=1}^{T}\left[\widehat{\mathbf{g}}_t\right]_i^2}\leq 2G_\infty\sqrt{T}.
		\end{split}
	\end{equation}
	By substituting \eqref{vt12} and \eqref{gt12} into \eqref{regret00}, \eqref{regret01} is obtained. \hfill $\blacksquare$

	\textbf{Proposition \ref{ConverAMSG}} implies that $\lim_{T \to \infty} \frac{R_T}{T} = 0$. 
%	\begin{equation}
%		\lim_{T \to \infty} \frac{R_T}{T} = 0.
%	\end{equation}
	This indicates that the proposed algorithm achieves a ‘no-regret’ property. This vanishing average regret as $T$ approaches infinity rigorously guarantees the convergence of the proposed AMSGrad-based optimization scheme, ensuring that the iterative sequence of antenna positions effectively minimizes the objective function over iterations.
		
	\begin{algorithm}[!t]
		\caption{The proposed AMSGrad method for solving (\ref{prob_222})}
		\begin{algorithmic}[1]
			\STATE Initialize $\mathbf{t}_{n,(k,0)} =\mathbf{t}_{n,k} $.
			\STATE \textbf{Repeat}
			\STATE \hspace{0.5cm}Compute the gradient $\bm{\nabla}_{\mathbf{t}_n}$ via \eqref{grad_tn_minor}. 
			\STATE \hspace{0.5cm}Compute the first-order moment via \eqref{hat_m_t}.
			\STATE \hspace{0.5cm}Compute the second-order moment via \eqref{hat_v_t}.
			\STATE \hspace{0.5cm}Update $	\widehat{\mathbf{f}}_t$ via \eqref{f_2}.
			\STATE \hspace{0.5cm}Update $\mathbf{t}_{n}$ via solving \eqref{prob_2223}.
			\STATE \hspace{0.5cm} $t \gets t+1$.
			\STATE \textbf{Until} Convergence criterion is met. 
			\STATE \textbf{return} $\mathbf{t}_{n,k}=\mathbf{t}_{n,(k,t)}$.
		\end{algorithmic}
		\label{alg3}
	\end{algorithm}

	%The convergence of the AO-based method can be guaranteed by \cite{xie2022perceptive}.
	By alternatively updating $\mathbf{F}$ and $\mathbf{T}$, the variables will converge to a stationary point as $m$ increases. %Subsequently, the precoding matrix $\mathbf{F}$ can be derived by performing Singular Value Decomposition (SVD) on $\mathbf{\Theta}$.

	\section{Simulation Results}\label{Sim_Res}
In this section, we provide simulation results to show the effectiveness of the proposed AO-based method. In the following, we first introduce the system parameters, and the benchmark algorithms.

\textbf{System Parameters:} We consider a millimeter-wave (mmWave) system operating at a carrier frequency of 28~GHz. Unless specified otherwise, the transmitter and receiver are equipped with $N = 8$ and $M = 4$ antennas, respectively. The number of eavesdroppers is also set as $M = 4$.  The feasible region of MA is set as $D_x = D_y = 50\lambda$. The K-Rician factors are set as $K_{e,1} = K_{e,2} =\cdots = K_{e,M} = K_{e} = 4$. The path gain for the $l$-th path in the legitimate channel and $\mathrm{Eve}_m$ is modeled as $\beta_{b,l}, \beta_{e,m} \sim \mathcal{CN}(0, 10^{-0.1\vartheta(d)})$, where $$\vartheta(d) = a + 10b\log_{10}(d) + \epsilon,$$ with $d$ denoting the propagation distance and $\epsilon \sim \mathcal{CN}(0, \sigma_{\epsilon}^2)$ \cite{6834753}. 
Following \cite{6834753}, we set $a = 61.4$, $b = 2$, and $\sigma_{\epsilon} = 5.8$~dB. The distance between the transmitter and receiver is set as $40$ m.  
The number of paths is set as $L = 16$. 
We consider a single communication user and $M$ eavesdroppers whose angle of departure (AoD) and angle of arrival (AoA) are randomly generated from $[10^\circ,30^\circ]$ and $[40^\circ,70^\circ]$, respectively. The noise power is set to $\sigma^2 = -90$~dBm. 
The parameters of AMSGrad-based method are set as $\beta_1 = 0.9$, $\lambda_1 = 0.99$, $\beta_{2} = 0.9$, and $\alpha = 0.5$.

\textbf{Benchmark methods:} To evaluate the performance of the proposed methods, we consider the following benchmark schemes for antenna position and precoding designs:

\emph{1) Antenna Configuration:} We compare the proposed MA optimization against a conventional FPA setup, i.e., the ULA along $x$-axis with a  fixed half-wavelength inter-element spacing.

\emph{2) Precoding Design:} The proposed precoding design is compared with the widely-used zero-forcing (ZF) precoding scheme \cite{1603708} to demonstrate the gains achieved by our GP-based optimization framework. In particular, the ZF precoder is defined by
\begin{equation}
\mathbf{F}_{\mathrm{ZF}} =  \frac{\sqrt{P}}{\left\Vert \widetilde{\mathbf{F}}_{\mathrm{ZF}} \right\Vert} \widetilde{\mathbf{F}}_{\mathrm{ZF}},
\end{equation}
where 
\begin{equation} 
\begin{aligned}
&\widetilde{\mathbf{F}}_{\mathrm{ZF}} =  \left[\mathbf{H}_{\mathrm{all}} \left(\mathbf{H}_{\mathrm{all}}^\HH
\mathbf{H}_{\mathrm{all}}\right)^{-1}\right]_{:,1:M},\\
&\mathbf{H}_{\mathrm{all}} = \left[ \mathbf{H},\mathbf{G}_{\mathrm{LoS}} \mathbf{\Lambda}^{\frac{1}{2}} \right].
\end{aligned}
\end{equation}

\subsection{Validation of Proposition \ref{prop:ergodic_secrecyC}}

\begin{figure}[t]
	\centering
	\includegraphics[width=3.6in]{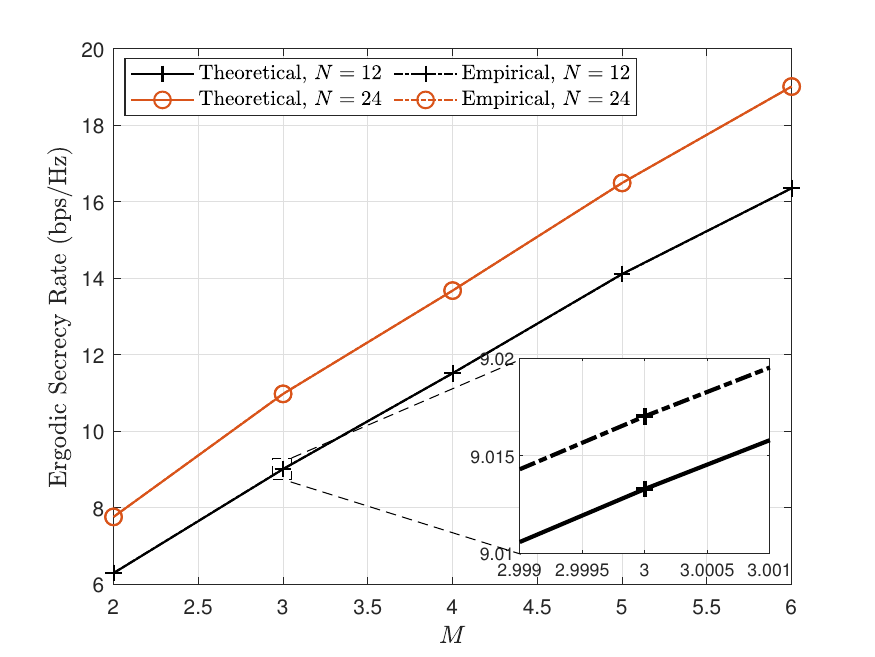}
	\caption{The ergodic secrecy rate  $\overline{\mathcal{R}}$ versus $M$.}
	\label{fig_accu_M}
\end{figure}

The legend `Empirical' represents the ESR performance obtained through Monte Carlo simulations, where each abscissa is averaged over 1000 independent trials. The legend `Theoretical' represents the deterministic equivalent for the ESR, which is defined in \eqref{App1}. 
As illustrated in Fig. \ref{fig_accu_M}, we investigate the accuracy of the deterministic approximation introduced in \textbf{\emph{Proposition \ref{prop:ergodic_secrecyC}}}. While the theoretical framework of \textbf{\emph{Proposition \ref{prop:ergodic_secrecyC}}} is rigorously predicated on the asymptotic regime where both the number of antennas $N$ and the number of eavesdroppers $M$ approach infinity, our numerical evaluations reveal a robust applicability to finite-dimensional systems. Specifically, the approximation demonstrates high precision even for moderately sized configurations (e.g., $M\geq 3$), significantly relaxing the constraints typically associated with large-system limits. Furthermore, the figure corroborates the law of RMT, as the approximation accuracy monotonically improves with an increasing number of snapshots. This trend confirms that the empirical results converge tightly to their deterministic equivalents as $M$ grows.

Moreover, a similar trend is observed regarding the number of transmit antennas $N$. As $N$  increases, the discrepancy between the analytical approximation and the Monte Carlo simulations will diminish. This behavior stems from the channel hardening effect inherent in large-scale antenna systems, where the random fluctuations of the channel matrix asymptotically stabilize around their deterministic limits. Consequently, the nearly perfect alignment between the empirical and theoretical curves across various system dimensions validates the accuracy of our theoretical results. This justifies utilizing the deterministic equivalent in \eqref{App1} as a computationally efficient objective function for the subsequent joint optimization framework, thereby eliminating the need for time-consuming stochastic averaging.

\begin{figure}[t]
	\centering
	\includegraphics[width=3.6in]{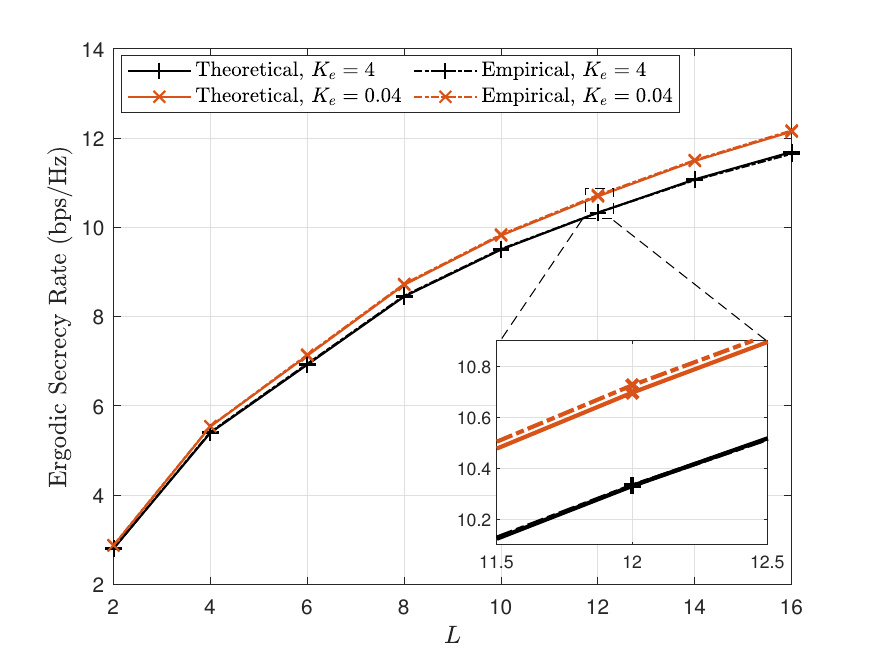}
	\caption{The ergodic secrecy rate  $\overline{\mathcal{R}}$ versus $L$.}
	\label{fig_accu_Ke}
\end{figure}

In Fig. \ref{fig_accu_Ke}, we investigate the impact of the number of paths in the legitimate channel $L$ and the Rician K-factor in the eavesdropper channel $K_e$. We set $N = 4$ and $M = 2$.
First, a close agreement between the analytical results and the Monte Carlo simulations is consistently observed across varying values of $L$, validating the robustness of our derived approximation in diverse scattering environments.
Furthermore, we examine the influence of the Rician factor $K_e$, which dictates the relative power dominance between the deterministic LoS component and the random NLoS scattering. Intuitively, a variation in $K_e$ significantly alters the statistical distribution of the channel. 
Crucially, Fig. \ref{fig_accu_Ke} demonstrates that the proposed deterministic equivalent \eqref{App1} maintains high accuracy across the entire range of $K_e$. Whether the channel is dominated by significant randomness (low $K_e$, large $L$) or strong deterministic components (high $K_e$), our asymptotic approximation effectively captures the system performance. This resilience to variations in channel statistics ($L$ and $K_e$) further confirms the suitability of using the derived deterministic equivalent as a reliable objective function for optimization in dynamic wireless environments.

\subsection{Convergence Performance}

\begin{figure}[t]
	\centering
	\includegraphics[width=3.6in]{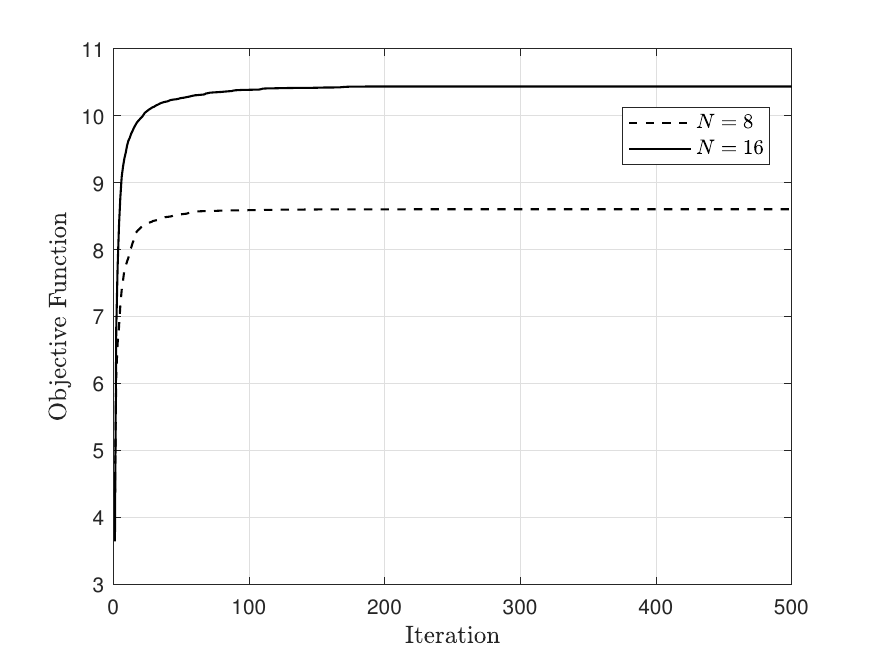}
	\caption{The convergence performance of Algorithm \ref{alg2}.}
	\label{fig_opt_F}
\end{figure}

Fig. \ref{fig_opt_F} illustrates the convergence behavior of the proposed GP-based algorithm, which is summarized in Algorithm \ref{alg2}. In this simulation, the number of paths in the legitimate channel is fixed at $L =4$. It can be observed that the algorithm exhibits monotonic convergence across various numbers of antennas $N$. Furthermore, the results indicate that the achievable secrecy performance improves significantly as $N$ increases. This enhancement is attributed to the additional spatial DoF provided by larger antenna arrays, which effectively bolster the system’s secrecy capacity.

\begin{figure}[t]
	\centering
	\includegraphics[width=3.6in]{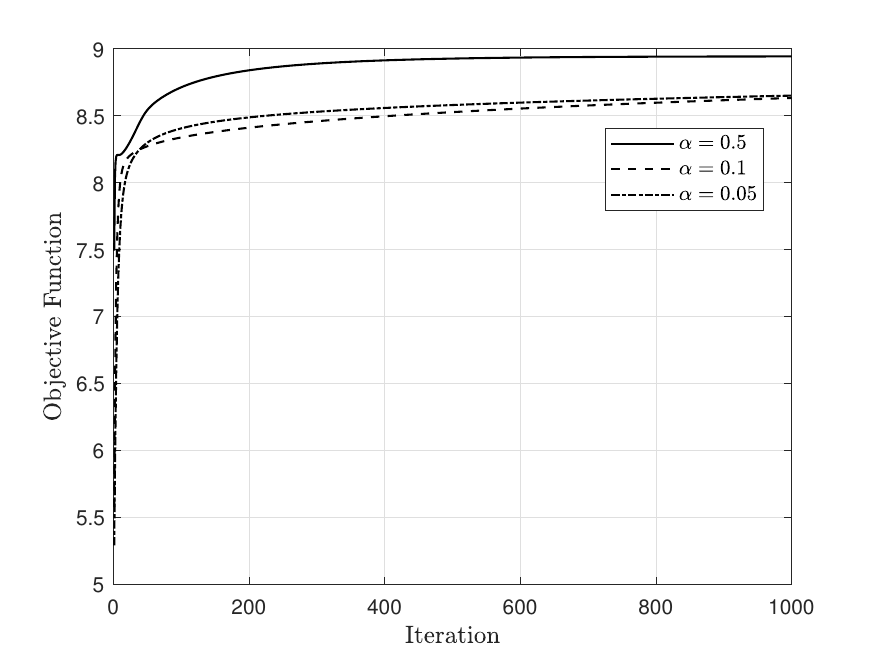}
	\caption{The convergence performance of Algorithm \ref{alg3}.}
	\label{fig_opt_T}
\end{figure}

Fig. \ref{fig_opt_T} illustrates the convergence trajectory of the proposed AMSGrad-based optimization framework, as detailed in Algorithm \ref{alg3}. The simulation setup assumes $N = 8$ MAs at Alice and $L = 4$ multipath components over the Alice-Bob channel. The results clearly demonstrate that the convergence rate is highly sensitive to the chosen step size schedule, $\alpha_t = \alpha/\sqrt{t}$. There is a distinct trade-off engaged here: as the scaling factor $\alpha$ decreases, the algorithm takes smaller steps in the gradient direction. Consequently, while potentially more stable, the convergence speed slows down significantly, requiring a higher computational budget in terms of iteration count to achieve the optimal value.

 \begin{figure}[t]
 	\centering
 	\includegraphics[width=3.6in]{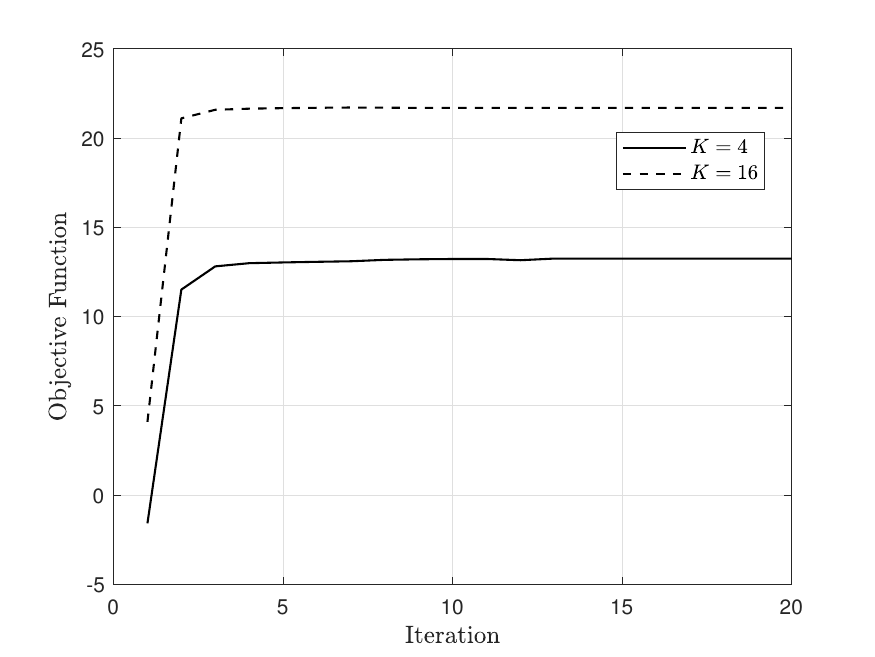}
 	\caption{The convergence performance of the AO-based framework.}
 	\label{fig_opt_TF}
 \end{figure}
 
 Fig. \ref{fig_opt_TF} illustrates the convergence behavior of the proposed AO-based framework. We can see that the algorithm exhibits rapid convergence, typically stabilizing within about 5 iterations across various channel settings.  Furthermore, the figure reveals a distinct performance improvement as the number of paths in the legitimate user’s channel, i.e., $L$, increases. This trend can be attributed to the enhanced spatial degrees of freedom available in richer scattering environments. Specifically, a larger $L$ contributes to a higher multiplexing gain, allowing the transmitter to exploit more spatial diversity to focus energy towards the legitimate user while nulling the eavesdropper. Consequently, the ESR performance is improved with the increase of $L$.

\subsection{ESR Performance}

\begin{figure}[t]
	\centering
	\includegraphics[width=3.6in]{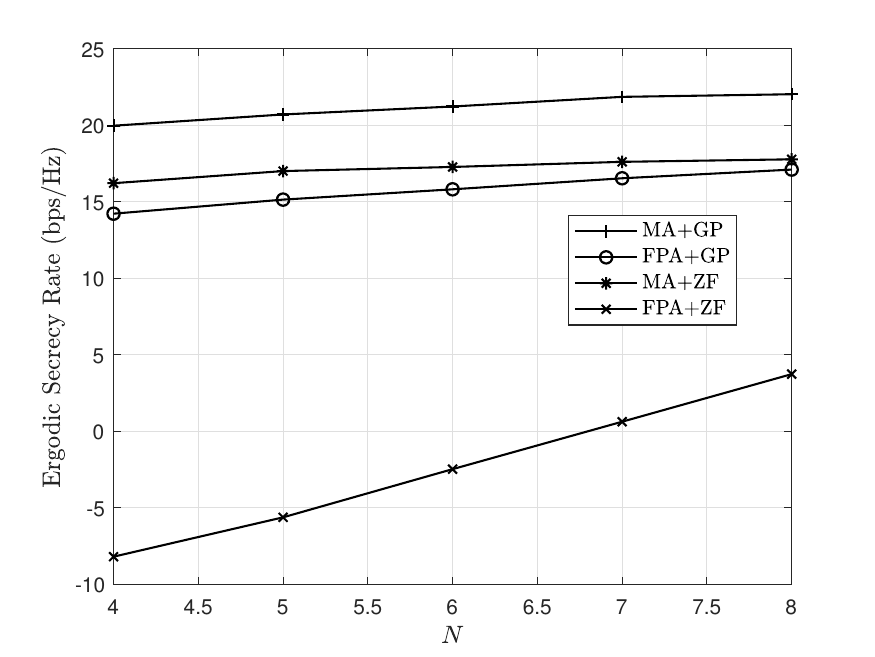}
	\caption{The ergodic secrecy rate  $\overline{\mathcal{R}}$ versus $N$.}
	\label{fig_overall_N}
\end{figure}

Fig. \ref{fig_overall_N} presents the ESR as a function of $N$. A monotonic increase in ESR is observed across all considered schemes as $N$ grows. This trend is expected, as a larger antenna array provides higher spatial DoF, enabling more precise beamforming to concentrate signal power on the legitimate receiver while simultaneously suppressing leakage to the eavesdropper. Notably, the proposed scheme, which integrates MAs with the GP-based optimization, consistently outperforms all baseline candidates. First, it can be observed that the ESR of the `FPA + ZF' scheme drops below $0$ bps/Hz. In the context of physical layer security, this negative value implies that the eavesdropper’s channel quality exceeds that of Bob, rendering secure communication impossible. As compared with the `MA + ZF' baseline, our proposed method achieves a substantial performance gain of approximately 4 bps/Hz. This significant gap underscores the sub-optimality of ZF in this security context. While ZF focuses solely on nulling the LoS of Eves, our proposed optimized precoder balances channel capacity maximization and information leakage more effectively, which can make more effective use of the available spatial resources. As compared with the `FPA + GP' benchmark, the proposed MA-aided scheme exhibits superior performance. Despite using the same advanced optimization algorithm, the fixed FPA is limited by its static geometry. In contrast, the introduction of MAs unlocks a new dimension of spatial DoF. By reconfiguring the antenna positions, the system can actively reshape the wireless channel to avoid deep fading for the legitimate user and degrade the eavesdropper’s channel quality, thereby expanding the achievable secrecy capacity region.

\begin{figure}[t]
	\centering
	\includegraphics[width=3.6in]{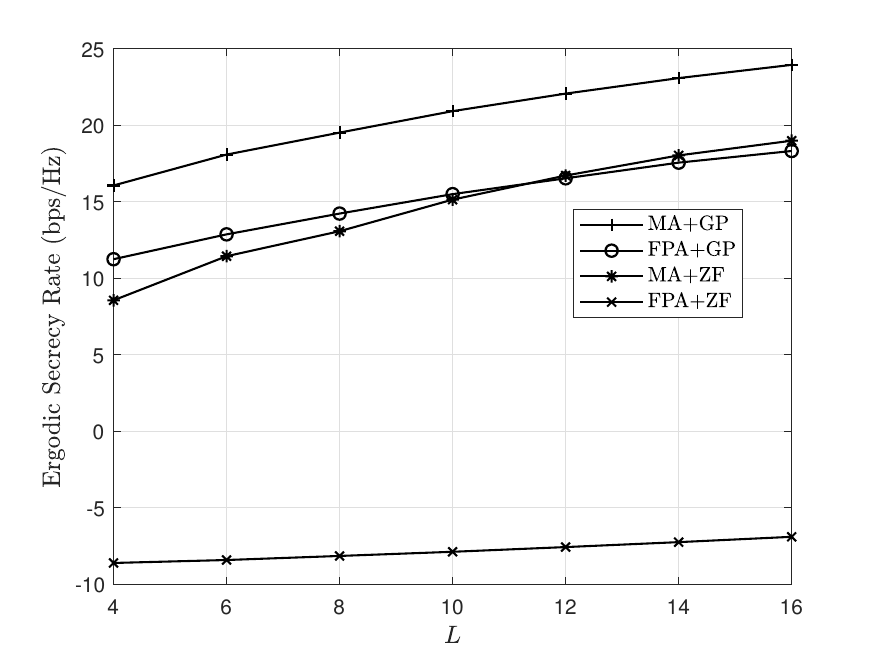}
	\caption{The ergodic secrecy rate  $\overline{\mathcal{R}}$ versus $L$.}
	\label{fig_overall_L}
\end{figure}

Fig. \ref{fig_overall_L} illustrates the impact of the number of paths, $L$, in Bob’s channel on the ESR. It can be observed that as $L$ increases, the performance gap between `MA+ZF' and `FPA+ZF' becomes more significant than that between `FPA+GP' and `FPA+ZF'. This trend suggests that the DoFs provided by antenna position optimization yield higher performance gains than precoding optimization in rich scattering environments. This phenomenon can be attributed to the nature of multipath propagation: a richer multipath environment induces more severe spatial fluctuations in the electromagnetic field, characterized by sharper peaks and deeper nulls within a confined region. Unlike FPAs, which may be constrained to locations with deep fading that precoding can only partially mitigate, MAs can exploit this increased spatial selectivity. By continuously adjusting their positions, MAs can capture local maxima of constructive interference, which become more pronounced and frequent as multipath components accumulate. Furthermore, while precoding is inherently limited by the fixed spatial correlation of a static array, MA position optimization proactively reshapes the channel matrix to enhance the orthogonality of channel vectors. Consequently, in an average sense, a larger $L$ increases the probability of finding antenna configurations that maximize the effective rank of the Alice-Bob channel.

	\section{Conclusion}

	In this paper, we have investigated secure communication within a MIMOME system enhanced by MAs, specifically addressing the challenge of imperfect ECSI. We adopted a practical channel model assuming access to instantaneous LoS and statistical NLoS components, employing the ESR as the performance metric. To overcome the analytical intractability of the exact ESR, we utilized large-dimensional RMT to derive a deterministic closed-form approximation. This analytical result not only facilitates efficient performance evaluation without Monte Carlo averaging but also provides the theoretical basis for system optimization. 	
	Guided by these derivations, we formulated a joint optimization problem to maximize the ESR by alternately designing the transmit precoding matrix and the MA positions. 
	Simulation results have corroborated the accuracy of our RMT-based approximation and demonstrated that the proposed MA-aided secure transmission scheme significantly outperforms conventional FPA benchmarks, highlighting the potential of MAs in enhancing physical layer security. 
	Moreover, although we considered statistical ECSI, the inevitable estimation errors in the LoS component or parameters of the statistical model (e.g., K-factor, correlation matrices) remain a concern. Future research could focus on robust designs that maximize the worst-case secrecy rate under bounded uncertainty sets for these parameters.



\begin{thebibliography}{10}
	\providecommand{\url}[1]{#1}
	\csname url@samestyle\endcsname
	\providecommand{\newblock}{\relax}
	\providecommand{\bibinfo}[2]{#2}
	\providecommand{\BIBentrySTDinterwordspacing}{\spaceskip=0pt\relax}
	\providecommand{\BIBentryALTinterwordstretchfactor}{4}
	\providecommand{\BIBentryALTinterwordspacing}{\spaceskip=\fontdimen2\font plus
		\BIBentryALTinterwordstretchfactor\fontdimen3\font minus
		\fontdimen4\font\relax}
	\providecommand{\BIBforeignlanguage}[2]{{%
			\expandafter\ifx\csname l@#1\endcsname\relax
			\typeout{** WARNING: IEEEtran.bst: No hyphenation pattern has been}%
			\typeout{** loaded for the language `#1'. Using the pattern for}%
			\typeout{** the default language instead.}%
			\else
			\language=\csname l@#1\endcsname
			\fi
			#2}}
	\providecommand{\BIBdecl}{\relax}
	\BIBdecl
	
	\bibitem{liu2022integrated}
	F.~Liu, Y.~Cui, C.~Masouros, J.~Xu, T.~X. Han, Y.~C. Eldar, and S.~Buzzi,
	``Integrated sensing and communications: Toward dual-functional wireless
	networks for {6G} and beyond,'' \emph{IEEE J. Sel. Areas Commun.}, vol.~40,
	no.~6, pp. 1728--1767, 2022.
	
	\bibitem{xie2023collaborative}
	L.~Xie, S.~Song, Y.~C. Eldar, and K.~B. Letaief, ``Collaborative sensing in
	perceptive mobile networks: Opportunities and challenges,'' \emph{IEEE
		Wireless Commun.}, vol.~30, no.~1, pp. 16--23, 2023.
	
	\bibitem{jiang2025network}
	Y.~Jiang, X.~Li, G.~Zhu, K.~Han, K.~Meng, W.~Yang, C.~Liu, Q.~Shi, and
	R.~Zhang, ``Network-{L}evel {P}erformance {A}nalysis for {A}ir-{G}round
	{I}ntegrated {S}ensing and {C}ommunication,'' \emph{IEEE Trans. Wireless
		Commun.}, 2025.
	
	\bibitem{xie2023networked}
	L.~Xie, S.~Song, and K.~B. Letaief, ``Networked {S}ensing {W}ith
	{AI}-{E}mpowered {I}nterference {M}anagement: {E}xploiting
	{M}acro-{D}iversity and {A}rray {G}ain in {P}erceptive {M}obile {N}etworks,''
	\emph{IEEE J. Sel. Areas Commun.}, vol.~41, no.~12, pp. 3863--3877, 2023.
	
	\bibitem{zhu2023modeling}
	L.~Zhu, W.~Ma, and R.~Zhang, ``Modeling and performance analysis for movable
	antenna enabled wireless communications,'' \emph{IEEE Trans. Wireless
		Commun.}, vol.~23, no.~6, pp. 6234--6250, 2023.
	
	\bibitem{ma2023mimo}
	W.~Ma, L.~Zhu, and R.~Zhang, ``Mimo capacity characterization for movable
	antenna systems,'' \emph{IEEE Trans. Wireless Commun.}, vol.~23, no.~4, pp.
	3392--3407, 2023.
	
	\bibitem{mei2024movable2}
	W.~Mei, X.~Wei, B.~Ning, Z.~Chen, and R.~Zhang, ``Movable-antenna position
	optimization: A graph-based approach,'' \emph{IEEE Wireless Commun. Lett.},
	vol.~13, no.~7, pp. 1853--1857, 2024.
	
	\bibitem{wang2025movable}
	D.~Wang, W.~Mei, B.~Ning, Z.~Chen, and R.~Zhang, ``Movable antenna enhanced
	wide-beam coverage: Joint antenna position and beamforming optimization,''
	\emph{IEEE Trans. Wireless Commun.}, 2025.
	
	\bibitem{wang2018survey}
	D.~Wang, B.~Bai, W.~Zhao, and Z.~Han, ``A survey of optimization approaches for
	wireless physical layer security,'' \emph{IEEE Commun. Surv. Tutor.},
	vol.~21, no.~2, pp. 1878--1911, 2018.
	
	\bibitem{xie2023sensing}
	L.~Xie, F.~Liu, J.~Luo, and S.~Song, ``Sensing {M}utual {I}nformation {W}ith
	{R}andom {S}ignals in {G}aussian {C}hannels,'' \emph{IEEE Trans. Commun.},
	vol.~73, no.~10, pp. 9437--9452, 2025.
	
	\bibitem{wei2022toward}
	Z.~Wei, F.~Liu, C.~Masouros, N.~Su, and A.~P. Petropulu, ``Toward
	multi-functional {6G} wireless networks: Integrating sensing, communication,
	and security,'' \emph{IEEE Commun. Mag.}, vol.~60, no.~4, pp. 65--71, 2022.
	
	\bibitem{kihero20236g}
	A.~B. Kihero, H.~M. Furqan, M.~Sahin, and H.~Arslan, ``{6G} and beyond wireless
	channel characteristics for physical layer security: Opportunities and
	challenges,'' \emph{IEEE Wireless Commun.}, vol.~31, no.~3, pp. 295--301,
	2023.
	
	\bibitem{10608156}
	X.~Zhu, J.~Liu, L.~Lu, T.~Zhang, T.~Qiu, C.~Wang, and Y.~Liu, ``Enabling
	{I}ntelligent {C}onnectivity: A {S}urvey of {S}ecure {ISAC} in {6G}
	{N}etworks,'' \emph{IEEE Commun. Surv. Tutor.}, vol.~27, no.~2, pp. 748--781,
	2025.
	
	\bibitem{11004012}
	J.~Wang, H.~Du, Y.~Liu, G.~Sun, D.~Niyato, S.~Mao, D.~In~Kim, and X.~Shen,
	``Generative {AI} {B}ased {S}ecure {W}ireless {S}ensing for {ISAC}
	{N}etworks,'' \emph{IEEE Trans. Inf. Forensics Security}, vol.~20, pp.
	5195--5210, 2025.
	
	\bibitem{5306434}
	X.~Zhou and M.~R. McKay, ``Physical layer security with artificial noise:
	Secrecy capacity and optimal power allocation,'' in \emph{2009 3rd
		International Conference on Signal Processing and Communication Systems},
	2009, pp. 1--5.
	
	\bibitem{6094170}
	N.~Romero-Zurita, M.~Ghogho, and D.~McLernon, ``Outage {P}robability {B}ased
	{P}ower {D}istribution {B}etween {D}ata and {A}rtificial {N}oise for
	{P}hysical {L}ayer {S}ecurity,'' \emph{IEEE Signal Process. Lett.}, vol.~19,
	no.~2, pp. 71--74, 2012.
	
	\bibitem{zheng2010optimal}
	G.~Zheng, L.-C. Choo, and K.-K. Wong, ``Optimal cooperative jamming to enhance
	physical layer security using relays,'' \emph{IEEE Trans. Signal Process.},
	vol.~59, no.~3, pp. 1317--1322, 2010.
	
	\bibitem{hu2017cooperative}
	L.~Hu, H.~Wen, B.~Wu, F.~Pan, R.-F. Liao, H.~Song, J.~Tang, and X.~Wang,
	``Cooperative jamming for physical layer security enhancement in {I}nternet
	of {T}hings,'' \emph{IEEE Internet Things J.}, vol.~5, no.~1, pp. 219--228,
	2017.
	
	\bibitem{yang2012cooperative}
	Y.~Yang, Q.~Li, W.-K. Ma, J.~Ge, and P.~Ching, ``Cooperative secure beamforming
	for {AF} relay networks with multiple eavesdroppers,'' \emph{IEEE Signal
		Process. Lett.}, vol.~20, no.~1, pp. 35--38, 2012.
	
	\bibitem{asaad2022secure}
	S.~Asaad, Y.~Wu, A.~Bereyhi, R.~R. Mueller, R.~F. Schaefer, and H.~V. Poor,
	``Secure active and passive beamforming in {IRS}-aided {MIMO} systems,''
	\emph{IEEE Trans. Inf. Forensics Security}, vol.~17, pp. 1300--1315, 2022.
	
	\bibitem{cheng2024enabling}
	Z.~Cheng, N.~Li, J.~Zhu, X.~She, C.~Ouyang, and P.~Chen, ``Enabling secure
	wireless communications via movable antennas,'' in \emph{ICASSP 2024-2024
		IEEE International Conference on Acoustics, Speech and Signal Processing
		(ICASSP)}.\hskip 1em plus 0.5em minus 0.4em\relax IEEE, 2024, pp. 9186--9190.
	
	\bibitem{mei2024movable}
	W.~Mei, X.~Wei, Y.~Liu, B.~Ning, and Z.~Chen, ``Movable-antenna position
	optimization for physical-layer security via discrete sampling,'' in
	\emph{GLOBECOM 2024-2024 IEEE Global Communications Conference}.\hskip 1em
	plus 0.5em minus 0.4em\relax IEEE, 2024, pp. 4750--4755.
	
	\bibitem{tang2024secure}
	J.~Tang, C.~Pan, Y.~Zhang, H.~Ren, and K.~Wang, ``Secure {MIMO} communication
	relying on movable antennas,'' \emph{IEEE Trans. Commun.}, 2024.
	
	\bibitem{shen2025movable}
	X.~Shen, X.~Wei, W.~Mei, Z.~Chen, J.~Fang, and B.~Ning,
	``Movable-antenna-enhanced physical-layer service integration: Performance
	analysis and optimization,'' \emph{IEEE Wireless Commun. Lett.}, 2025.
	
	\bibitem{li2019beamforming}
	Q.~Li and L.~Yang, ``Beamforming for cooperative secure transmission in
	cognitive two-way relay networks,'' \emph{IEEE Trans. Inf. Forensics
		Security}, vol.~15, pp. 130--143, 2019.
	
	\bibitem{feng2024movable}
	Z.~Feng, Y.~Zhao, K.~Yu, and D.~Li, ``Movable antenna empowered physical layer
	security without eve's csi: Joint optimization of beamforming and antenna
	positions,'' \emph{arXiv preprint arXiv:2405.16062}, 2024.
	
	\bibitem{hu2024secure}
	G.~Hu, Q.~Wu, K.~Xu, J.~Si, and N.~Al-Dhahir, ``Secure wireless communication
	via movable-antenna array,'' \emph{IEEE Signal Process. Lett.}, vol.~31, pp.
	516--520, 2024.
	
	\bibitem{11214460}
	K.~Li, K.~Yu, D.~Ma, Y.~Zhao, X.~Liu, Q.~Zhang, and Z.~Feng, ``Can {M}ovable
	{A}ntenna-enabled {M}icro-{M}obility {R}eplace {UAV}-enabled
	{M}acro-{M}obility? a {P}hysical {L}ayer {S}ecurity {P}erspective,''
	\emph{IEEE Trans. Mob. Comput.}, pp. 1--13, 2025.
	
	\bibitem{su2023sensing}
	N.~Su, F.~Liu, and C.~Masouros, ``Sensing-assisted eavesdropper estimation: An
	{ISAC} breakthrough in physical layer security,'' \emph{IEEE Trans. Wireless
		Commun.}, vol.~23, no.~4, pp. 3162--3174, 2023.
	
	\bibitem{hu2024movable}
	G.~Hu, Q.~Wu, D.~Xu, K.~Xu, J.~Si, Y.~Cai, and N.~Al-Dhahir, ``Movable
	{A}ntennas-{A}ssisted {S}ecure {T}ransmission {W}ithout {E}avesdroppers’
	{I}nstantaneous {CSI},'' \emph{IEEE Trans. Mob. Comput.}, vol.~23, no.~12,
	pp. 14\,263--14\,279, 2024.
	
	\bibitem{kang2006capacity}
	M.~Kang and M.-S. Alouini, ``Capacity of correlated {MIMO} {R}ayleigh
	channels,'' \emph{IEEE Trans. Wireless Commun.}, vol.~5, no.~1, pp. 143--155,
	2006.
	
	\bibitem{5429113}
	J.~Dumont, W.~Hachem, S.~Lasaulce, P.~Loubaton, and J.~Najim, ``On the
	{C}apacity {A}chieving {C}ovariance {M}atrix for {R}ician {MIMO} {C}hannels:
	{A}n {A}symptotic {A}pproach,'' \emph{IEEE Trans. Inf. Theory}, vol.~56,
	no.~3, pp. 1048--1069, 2010.
	
	\bibitem{xie2022perceptive}
	L.~Xie, P.~Wang, S.~Song, and K.~B. Letaief, ``Perceptive mobile network with
	distributed target monitoring terminals: Leaking communication energy for
	sensing,'' \emph{IEEE Trans. Wireless Commun.}, vol.~21, no.~12, pp.
	10\,193--10\,207, 2022.
	
	\bibitem{sun2016majorization}
	Y.~Sun, P.~Babu, and D.~P. Palomar, ``Majorization-minimization algorithms in
	signal processing, communications, and machine learning,'' \emph{IEEE Trans.
		Signal Process.}, vol.~65, no.~3, pp. 794--816, 2016.
	
	\bibitem{absil2008optimization}
	P.-A. Absil, R.~Mahony, and R.~Sepulchre, \emph{Optimization algorithms on
		matrix manifolds}.\hskip 1em plus 0.5em minus 0.4em\relax Princeton
	University Press, 2008.
	
	\bibitem{reddi2019convergence}
	S.~J. Reddi, S.~Kale, and S.~Kumar, ``On the convergence of adam and beyond,''
	\emph{arXiv preprint arXiv:1904.09237}, 2019.
	
	\bibitem{cvx}
	M.~Grant and S.~Boyd, ``{CVX}: Matlab {S}oftware for {D}isciplined {C}onvex
	{P}rogramming, version 2.1,'' \url{http://cvxr.com/cvx}, Mar. 2014.
	
	\bibitem{6834753}
	M.~R. Akdeniz, Y.~Liu, M.~K. Samimi, S.~Sun, S.~Rangan, T.~S. Rappaport, and
	E.~Erkip, ``Millimeter {W}ave {C}hannel {M}odeling and {C}ellular {C}apacity
	{E}valuation,'' \emph{IEEE J. Sel. Areas Commun.}, vol.~32, no.~6, pp.
	1164--1179, 2014.
	
	\bibitem{1603708}
	T.~Yoo and A.~Goldsmith, ``On the optimality of multiantenna broadcast
	scheduling using zero-forcing beamforming,'' \emph{IEEE J. Sel. Areas
		Commun.}, vol.~24, no.~3, pp. 528--541, 2006.
	
\end{thebibliography}
\end{document}